\documentclass[11pt,a4paper]{article}
\usepackage{jcappub}
\usepackage{bm}
\usepackage{times}

\newcommand{\dN}{{d\bar N\over dz d\Omega}}
\newcommand{\DDD}{D}
\newcommand{\dd}{{\cal\DDD}}
\newcommand{\dD}{\de\DDD}
\newcommand{\Ddd}{\de\dd}
\newcommand{\DDd}{\De\dd}
\newcommand{\CCC}{\bdv{C}}
\newcommand{\KKK}{\bdv{K}}
\newcommand{\MMM}{\bdv{M}}
\newcommand{\pdf}{\mathcal{P}}
\newcommand{\II}{\tilde C}     
\newcommand{\IS}{\tilde S}     
\newcommand{\sss}{s}
\newcommand{\DM}{\bdv{D}}
\newcommand{\hdd}{\hat{\bar\dd}}
\newcommand{\GD}{\mathcal{G}}
\newcommand{\FF}{\mathcal{F}}
\newcommand{\NK}{\mathcal{N}}
\newcommand{\HK}{\mathcal{H}}
\newcommand{\up}[1]{{\rm #1}}
\newcommand{\bdv}[1]{{\bf #1}}
\newcommand{\beeq}{\begin{equation}}
\newcommand{\eneq}{\end{equation}}
\newcommand{\bear}{\begin{eqnarray}}
\newcommand{\enar}{\end{eqnarray}}
\newcommand{\nnn}{\nonumber \\}
\newcommand{\nn}{\nonumber}
\newcommand{\AVE}[1]{\left\langle#1\right\rangle}
\newcommand{\RA}{\rightarrow}
\newcommand{\pa}{\partial}
\newcommand{\Dquad}{\qquad\qquad}
\newcommand{\EQR}[1]{Eq.~\eqref{#1}}
\newcommand{\EQRS}[1]{Eqs.~\eqref{#1}}
\newcommand{\mpc}{{\rm Mpc}}
\newcommand{\hmpc}{{h^{-1}\mpc}}
\newcommand{\hmpci}{{h\mpc^{-1}}}
\newcommand{\OO}{\mathcal{O}}
\newcommand{\De}{\Delta}
\newcommand{\xvec}{\bdv{x}}
\newcommand{\Vang}{\bdv{\hat n}}
\newcommand{\Tr}{\up{Tr}}
\newcommand{\ga}{\gamma}
\newcommand{\de}{\delta}
\newcommand{\Om}{\Omega}
\newcommand{\rbar}{\bar r}      
\newcommand{\dL}{\mathcal{D}_L}      
\newcommand{\ddL}{\delta\mathcal{D}_L}
\newcommand{\ttt}{\theta}          
\newcommand{\pp}{\phi}

\begin{document}

\begin{titlepage}

\setcounter{page}{1} \baselineskip=15.5pt \thispagestyle{empty}
\pagenumbering{roman}

\bigskip

\vspace{1cm}
\begin{center}
{\fontsize{20}{28}\selectfont \bfseries Cosmological Information Contents on
the Light-Cone}
\end{center}

\vspace{0.2cm}

\begin{center}
{\fontsize{13}{30}\selectfont Jaiyul Yoo,$^{a,b}$ Ermis Mitsou,$^a$
Nastassia Grimm,$^a$ Ruth Durrer$^c$ and Alexandre~Refregier$^d$}
\end{center}

\begin{center}
\vskip 8pt
\textsl{$^a$ Center for Theoretical Astrophysics and Cosmology,
Institute for Computational Science}\\
\textsl{University of Z\"urich, Winterthurerstrasse 190,
CH-8057, Z\"urich, Switzerland}

\vskip 7pt

\textsl{$^b$Physics Institute, University of Z\"urich,
Winterthurerstrasse 190, CH-8057, Z\"urich, Switzerland}

\vskip 7pt

\textsl{$^c$D\'epartement de Physique Th{\'e}orique \& Center 
for Astroparticle Physics, Universit\'e de Gen\`eve\\
Quai E. Ansermet 24, CH-1211 Gen\`eve 4, Switzerland}

\vskip 7pt

\textsl{$^d$Institute for Particle Physics and Astrophysics, Department
of Physics, ETH Z\"urich, Wolfgang-Pauli-Strasse 27, CH-8057 Z\"urich,
Switzerland}

\vskip 7pt

\today

\end{center}

\note{jyoo@physik.uzh.ch,~~ ermitsou@physik.uzh.ch,~~ ngrimm@physik.uzh.ch,
~~ruth.durrer@unige.ch,~~
alexandre.refregier@phys.ethz.ch}

\vspace{1.2cm}
\hrule \vspace{0.3cm}
\noindent {\sffamily \bfseries Abstract} \\[0.1cm]
We develop a theoretical framework to describe the cosmological observables
on the past light cone such as the luminosity distance, weak lensing, 
galaxy clustering, and the cosmic microwave background 
anisotropies. We consider that all the cosmological observables include
not only the background quantity, but also the perturbation quantity, and
they are subject to cosmic variance, which sets the fundamental limits
on the cosmological information that can be derived from such observables,
even in an idealized survey with an infinite number of observations.
To quantify the maximum cosmological information content, we
apply the Fisher information matrix formalism and spherical harmonic
analysis to  cosmological observations,
in which the angular and the radial positions of the observables on the light
cone carry different information. We discuss the maximum cosmological 
information that can be derived
from five different observables:
(1)~type Ia supernovae, (2)~cosmic microwave background anisotropies,
(3)~weak gravitational lensing, 
(4)~local baryon density, and (5)~galaxy clustering.
We compare our results with the cosmic variance obtained in the standard
approaches, which treat the light cone volume as a cubic box of simultaneity.
We discuss  implications of our formalism and ways to overcome
the fundamental limit.
\vskip 10pt
\hrule

\vspace{0.6cm}
\end{titlepage}

\noindent\hrulefill \tableofcontents \noindent\hrulefill

\pagenumbering{arabic}

\section{Introduction}
The perturbations in cosmological 
large-scale structure are generated by the quantum
fluctuations in the early Universe. On large scales, they are almost
fully characterized by a Gaussian distribution and its power spectrum.
Due to the random nature of the perturbations,
many independent modes need to be sampled to obtain a
robust estimate of the underlying power spectrum. Given a survey volume,
however, there exists a finite number of 
independent modes available for  large-scale
observations, and  these observations are subject to the large-scale
fluctuations, known as  sample variance. When the survey volume is
limited by our observable Universe, sample variance 
is called  cosmic variance, and it sets a limit to the fundamental cosmological 
information contents
available to us (see, e.g., \cite{PEEBL80,PEACO99,DODEL03,DURRE08}).
In the past, when the large-scale observations were limited to a small
sky coverage and redshift range, the survey volume was well approximated as
a cubic box of simultaneity, far away from the observer, so that the 
traditional 3D Fourier analysis provided a useful way to estimate the
cosmological information contents in the survey. As we assume statistical homogeneity, each Fourier mode in a
rectangular volume is independent, and its sample variance is set by its own
power spectrum \cite{FEKAPE94,TEGMA97}. However,  experimental
and observational techniques have developed 
rapidly in recent years and the angular
coverage and redshift depth of large-scale surveys has become wider and
deeper, so that this simple approximation of estimating the cosmic variance
needs to be revisited.

Cosmological observables are mapped on the observer sky, always in terms of
 angular positions on the sky and redshifts. In particular, the radial
position obtained by the {\it observed} redshift carries different information
--- the observers locate the cosmological observables along the past light 
cone with the redshift, while the angular position of the cosmological
observables spans the two-dimensional sphere of  constant redshift
seen by the observer. Therefore, it is evident
that the traditional Fourier analysis in a rectangular box is fundamentally
incompatible with how the cosmological observables are mapped 
in the observer sky.
This inadequacy is maximally manifest in the analysis of the cosmic microwave
background (CMB) anisotropies, where the sky coverage is (almost) a full
sphere and the angular positions of the CMB
temperature and polarization anisotropies are decomposed in terms of
spherical harmonics with discrete angular momentum, rather than with continuous
Fourier modes. Moreover, significant progress has been  made in
the large-scale galaxy surveys to map the three-dimensional distribution of
the matter density, greatly improving upon the first-generation surveys
with the volume $\ll0.1$~Gpc$^3$.
In particular, the upcoming stage-IV surveys such as the
Dark Energy Spectroscopic Instrument \cite{DESI13} and
Large Synoptic Survey Telescope \cite{LSST04} and two space missions,
Euclid \cite{EUCLID11} and the Wide Field Infrared Survey Telescope 
\cite{WFIRST12}, will measure millions of galaxies with great precision
by observing together
almost a half of the entire sky over a large range of redshift.
In this era of precision cosmology, the traditional Fourier analysis
is increasingly inaccurate and becomes the source of systematic errors.

On the observed light cone we assume statistical isotropy so that the spherical harmonic modes are independent. However, the radial direction on the lightcone is  mixed with time evolution which breaks translation symmetry. Therefore large radial modes are not statistically independent and we expect cross correlation between different redshifts. Depending on the observable considered, these can be very relevant and contain important cosmological information.

Here, we develop a theoretical framework to generally
describe  cosmological
observables on the light cone and use the Fisher information matrix
to quantify the maximum cosmological information contents available from
observations of such cosmological observables. For 
two-dimensional angular observables such as  CMB anisotropies,
the standard formalism based on spherical harmonics
is as accurate as our new formalism, except for one subtlety
that  background quantities cannot be measured by observations from a
single light cone due to the perturbation of the 
monopole. This subtlety is often
ignored, leading to
an interesting bias and information loss, as we detail in 
section~\ref{ssec:CMB}. For  three-dimensional observables,
the spherical Fourier analysis based on the radial and angular eigenfunctions
of the Helmholtz equation is well developed \cite{BIQU91,FILAET95,HETA95},
and it has been applied to the observational data analysis 
\cite{FISCLA94,TABAET99} and to theoretical
predictions \cite{RARE12,SHCRPE12,LERAET12,YODE13,NIREET14,LARAST15}.
Complementing the Fourier analysis to the spherical harmonics decomposition,
the spherical Fourier analysis provides the most natural way to analyze
the cosmological observables in the observer sky. However, the difficulty
lies in computing the inverse spherical power spectrum, needed to quantify
 cosmic variance.
As opposed to the angular analysis or the traditional Fourier analysis,
where the inverse can be trivially obtained, the inverse spherical power
spectrum has not been derived in the spherical Fourier analysis.
Here we derive the inverse spherical power spectrum and use it to compute
the maximum cosmological information contents under the assumption of
Gaussianity. 
A similar idea was pursued \cite{LOEB12} to compute the cosmological
information contents as a function of the cosmic time, though, under the
simplifying assumptions using the traditional Fourier analysis.

The organization of the paper is as follows: We first develop a unified
theoretical framework for modeling the three-dimensional cosmological
observables, the angular observables, and the projected observables
in section~\ref{sec:observable}. In section~\ref{sec:like}, we derive the
likelihood of the cosmological observables on the light cone under the
assumption of Gaussianity. We then use the Fisher information technique
in section~\ref{sec:info} to quantify the maximum cosmological information
contents available from the cosmological observables. In section~\ref{sec:app},
we apply our Fisher matrix calculations to the luminosity distance,
the CMB anisotropies, 3D weak lensing, the cosmic baryon density measurements,
and the galaxy power spectrum. We discuss the implications of our new formulation of 
the cosmic variance on the light cone  in section~\ref{sec:discuss}. In 
Appendix~\ref{app:sp}, we present an alternative to the usual
spherical Fourier analysis, in which the radial position is decomposed
directly in terms of the observed redshift, avoiding the need to rely
on a fiducial cosmology. In Appendix~\ref{app:ob}, we present the relation
of the spherical power spectrum on the light cone 
to the usual power spectrum on a hypersurface of simultaneity.

\section{Cosmological Observables on the Light Cone}
\label{sec:observable}
Here we present the theoretical descriptions of  cosmological observables
on the light cone. We start with the most important observable, namely,
 number counts of luminous objects, which we collectively call galaxies.
Next, we provide
theoretical descriptions of other cosmological observables derived from
observations of galaxies such as the luminosity distance, the weak
lensing shear, and so on. Drawing on these theoretical descriptions
of cosmological observables, we consider two additional
 cases, in which the observations are limited to a single 
redshift bin and the observations 
are projected along the line-of-sight direction.

The main point of  our formalism is to account for the fact that the
cosmological observables are obtained only on the light cone volume, rather
than on a hypersurface of simultaneity. 
Compared to the standard descriptions, this consideration significantly changes
our theoretical descriptions of the cosmological observables in this
section and more
dramatically the cosmological information contents in section~\ref{sec:info}.

\subsection{Galaxy clustering and number counts}
\label{ssec:galaxy}
Luminous objects such as galaxies are easy to observe up to very high 
redshift, providing a great opportunity for us to probe the Universe.
In particular, their number density is the primary observable and its
two-point correlation (or the power spectrum) has been widely used to
test our theoretical models and to understand the Universe
(see, e.g., \cite{WEMOET13} for review). Galaxy clustering contains
a wealth of cosmological information. Its intrinsic correlation
encodes the underlying matter distribution, and the volume effects
involve the redshift space distortion and  gravitational lensing 
(see, e.g., \cite{KAISE84,KAISE87,YOO09}),
in addition to subtle relativistic effects.

In observations, we find a luminous object given the conditions for its color
and morphology and the thresholds for its brightness in the simplest case,
and the object that satisfies the conditions is identified as a galaxy
within a redshift bin $(z,z+dz)$  
and a solid angle $d\Omega=\sin\ttt d\ttt d\pp$.
Its position is then recorded in terms of 
the redshift~$z$ and angular position 
$\Vang=(\ttt,\pp)$, which we represent by~$\xvec_i$ for $i$-th 
galaxy. For its theoretical description, the observers often  
assume a cosmological model to  convert it in the
observer frame:
\beeq
\xvec:=\rbar(z)\Vang
=\rbar(z)(\sin\ttt\cos\pp,\sin\ttt\sin\pp,\cos\ttt)~.
\eneq
Throughout the paper,
we often use $\xvec=(z,\Vang)$ for notational simplicity.
The total number~$N$ of the observed galaxies,
\beeq
N=\sum_{i=1}^ N dN(\xvec_i)~,\Dquad dN\in\{0,1\}~,
\eneq
is simply the sum of the number~$dN$ of galaxies in a small 
volume~$d\bar V$ centered at~$\xvec_i$.
The number count~$dN(\xvec_i)$
is then further related to the (observed) galaxy number density~$n(\xvec_i)$
as
\beeq
dN(\xvec_i)={dN\over dzd\Omega}(\xvec_i)dzd\Omega:=n(\xvec_i)d\bar V(\xvec_i)~,
\eneq
where the (observed)  background volume element,
\beeq
d\bar V(z_i)={\rbar^2(z_i)\over H(z_i)(1+z_i)^3}~dz~d\Omega~,
\eneq
is the physical volume, corresponding to the observed redshift bin~$dz$ and the
solid angle~$d\Omega$ in an assumed homogeneous background universe. Given a set of cosmological
parameters, the observers can convert the redshift and angle to physical
distances by using the Hubble parameter~$H(z)$ and the angular diameter
distance
\beeq
\rbar(z)=\int_0^z{dz'\over H(z')}~,
\eneq
where we assume a flat universe ($K=0$).

To model the observations, we develop theoretical predictions that describe
such observations of galaxies at any (continuous) point~$\xvec$ 
within the survey volume (instead of discrete observation points~$\xvec_i$),
while keeping the total number~$N$ of the observed galaxies.
The observed galaxy number density is now modeled
as a continuous number density field
\beeq
\label{eq:gald}
n(\xvec):=\bar n(z)\left[1+\de_g(\xvec)\right]~,
\eneq
where we split the number density into the number density $\bar n(z)$ 
in the background and its perturbation~$\de_g$ 
around the background. Furthermore,
since the inhomogeneities in the Universe affect our cosmological observables,
the observed galaxy number density $n(\xvec)$ and the volume element $d\bar V$
are different from the physical galaxy number density~$n_p$ and the 
physical volume~$dV_p$ in the inhomogeneous Universe
that corresponds
to the observed position~$\xvec=(z,\Vang)$. 
Indeed, the observed number count is  related as
\beeq
\label{eq:dN}
dN(\xvec)=n_p(\xvec)dV_p(\xvec):=n(\xvec)d\bar V(\xvec)~,
\eneq
and these physical quantities can be further split as
\beeq
n_p(\xvec):=\bar n(z)\left[1+\de_s(\xvec)\right]~,\Dquad 
dV_p(\xvec):=d\bar V(\xvec)\left[1+\de V(\xvec)\right]~,
\eneq
where $\de_s$ is the intrinsic fluctuation of the galaxy number density and
$\de V$ is the (dimensionless) fluctuation of the physical volume element.
From \EQR{eq:dN}, the observed galaxy (number density) fluctuation is then 
derived as
\beeq
\de_g(\xvec)=\de_s(\xvec)+\de V(\xvec)+\de_s(\xvec) \de V(\xvec)~,
\eneq
and proper relativistic computations of the gauge-invariant 
expression~$\de_g$ and its correlations
have been the focus of recent research 
(see, e.g., \cite{YOFIZA09,YOO10,BODU11,CHLE11,JESCHI12,YOO14a,YOO14b,
TABOET18,SCYOBI18} for details),
as it involves subtle gauge issues and they can be used to extract
extra cosmological information. For our present purposes, we do not
need to know their detailed expressions, but it suffices to note two things:
(1)~Though the above expressions are exact, the individual
quantities are perturbations, such that the quadratic terms can be dropped
for the linear-order calculations. (2)~The individual terms are gauge-invariant
and expressed at the observed position~$\xvec$.

Before we proceed, we define a few more perturbation quantities associated
with the observed number counts. First, we define the background redshift
distribution
\beeq
\dN(z):={\bar n(z)\rbar^2(z)\over H(z)(1+z)^3}~,
\eneq
which is the number of galaxies we would measure 
per redshift bin~$dz$ and solid angle~$d\Omega$ in a homogeneous universe.
The observed number count is then modeled as a continuous field
\beeq
dN(\xvec)=\bar n(z)d\bar V(\xvec)\left[1+\de_g(\xvec)\right]
=\dN(z)dzd\Omega\left[1+\de_g(\xvec)\right]~,
\eneq
and the total number of the observed galaxies is simply
\bear
N(z_{\max}) &=&\int dN=\int_0^{z_{\max}}dz~\dN\int d\Omega \left(1+\de_g\right) \nonumber\\
&:=&
\bar N(z_{\max})+4\pi\int_0^{z_{\max}} dz~\dN~\AVE{\de_g}_\Om:=\bar N\left(1+\de N\right)~,
\enar
where we defined the background total number~$\bar N$ and its dimensionless
fluctuation~$\de N$ (both of them are independent of direction)
\beeq
\bar N(z_{\max}):=4\pi\int_0^{z_{\max} }dz~\dN~,\Dquad 
\de N(z_{\max}):=4\pi\int_0^{z_{\max}} dz~{1\over\bar N}\dN~\AVE{\de_g}_\Om~,
\eneq
and the angle-averaged galaxy fluctuation (or the monopole)
\beeq
\AVE{\de_g}_\Om(z):=\int{d\Omega\over4\pi}~\de_g(\xvec)=
\int{d\Omega\over4\pi}\left[\de_s(\xvec)+\de V(\xvec)+\de_s(\xvec) 
\de V(\xvec)\right]~.
\eneq
Note that only the total number~$N$ of the observed galaxies is a physical
number
and the split of~$N$ into~$\bar N$ and~$\de N$ is purely theoretical
for later convenience. Here we assume a full sky coverage for simplicity.

\subsection{Other cosmological observables}
\label{ssec:otherobs}
Beyond the primary observable or the galaxy number counts, other cosmological
information can be extracted from the observations of luminous objects
such as the luminosity distance from type~Ia supernovae, the lensing
shear from the shape of galaxies, and so on. These cosmological observables
can be used to compute their two-point correlation (or higher statistics)
in the  same way galaxy clustering is measured, and they contain equally
important cosmological information, compared to measurements of
galaxy clustering.

In addition to the galaxy number counts, we consider an observable
quantity~$\DDD$  from galaxies such as the luminosity distance
and construct the observed data set~$\dd_i^\up{obs}$
for $i$-th galaxy at~$\xvec_i$:
\beeq
\label{eq:dataset}
\bm{\dd}^\up{obs}=\{\DDD(\xvec_1),~\DDD(\xvec_2),~\cdots,~\DDD(\xvec_N)\}~,
\eneq
where we used the boldface letter to indicate the observed data set is a 
vector.
As we are often interested in the background quantity~$\bar\DDD(z)$ at
a given redshift~$z$, we may average the observed data set to derive an
estimate of~$\bar\DDD(z)$, if all the measurements
are in the same redshift bin:
\beeq
\label{eq:dataavg}
\left\langle\dd\right\rangle^\up{obs}(z)
:=\frac1N\sum_{i=1}^N\dd_i^\up{obs}~,\Dquad z_i\in(z,z+dz)~,
\eneq
where we define the notation for the average~$\AVE{\cdots}$ of observed
quantities.
In the same spirit, we model these observations and develop theoretical
predictions that provide the observable~$\DDD(\xvec)$ from galaxies at any 
(continuous) point~$\xvec$ (instead of discrete observation 
points~$\xvec_i$), and the observable quantity~$\DDD$ is
then approximated as a continuous field
\beeq
\label{eq:DDD}
\DDD(\xvec):=\bar\DDD(z)\left[1+\dD(\xvec)\right]~,
\eneq
where we again
split the observable quantity~$\DDD$ into a background~$\bar\DDD$
and the dimensionless
perturbation~$\dD$ around the background. The perturbation~$\dD$
is a diffeomorphism invariant scalar, and it is gauge-invariant at the 
linear order \cite{YODU17}. For example, the perturbation~$\dD$ for the
luminosity distance as an observable quantity~$\DDD$
has been computed (see, e.g., \cite{BODUGA06,BODUKU06,YOSC16,BIYO16,SCYO17,
BIYO17}), and it involves the
Doppler effects and subtle relativistic effects associated with the light
propagation. Here, we leave it general as the detailed expressions are
less important for our current discussion.

Note, however, that the continuous observable field~$\DDD(\xvec)$ alone
cannot be a complete description of our observed data set~$\dd_i^\up{obs}$ in
the continuous limit, as we measure the
observable~$\DDD$ only through observations of galaxies at~$\xvec_i$.
The observed data set~$\dd_i^\up{obs}$ in \EQR{eq:dataset} is not obtained
by uniform sampling of~$\DDD$ over the survey volume, but by
sampling biased objects such as galaxies. For example, considering the
luminosity distance observations from supernovae, we have fewer measurements
of the luminosity distance (or none) in an underdense region, where there are 
fewer galaxies and supernovae, despite the fact that the luminosity distance
to this underdense region is non-zero. The observed data set $\dd^\up{obs}_i$
in \EQR{eq:dataset} is indeed a set of observables
weighted by the galaxy number counts, such that our theoretical description is
\beeq
\label{eq:ddNN}
\dd(\xvec)=\DDD(\xvec){dN(\xvec)\over N}~,
\eneq
where $dN(\xvec)$ becomes zero or one in the observed data set
and we simply
divided by the (constant) total number~$N$ of galaxies for later convenience.
For instance, 
the angle average of~$\DDD(\xvec)$
does not reproduce the observed angle average in \EQR{eq:dataavg}, but
the angle average of~$\dd(\xvec)$ does, as in \EQR{eq:theavg}.
Finally, we split our model for the observed data set
\beeq
\label{eq:theory}
\dd(\xvec):=\bar\dd(z)\left[1+\Ddd(\xvec)\right]~,
\eneq
into the background and the perturbation parts 
\bear
\label{eq:bgddd}
\bar\dd(z)&=&{\bar\DDD(z)\over\bar N}\dN(z)dz d\Om
={\bar\DDD(z)\bar n(z)\rbar^2(z)\over \bar NH(z)(1+z)^3}dz d\Om
=:\hdd(z)dzd\Om~,\\
\label{eq:perddd}
\Ddd(\xvec)&=&{\left[1+\dD(\xvec)\right]\left[1+\de_g(\xvec)\right]
\over1+\de N}-1={\dD(\xvec)+\de_g(\xvec)-\de N+\dD(\xvec)\de_g(\xvec)
\over1+\de N}~.
\enar
We include the observational bin sizes for the redshift~$dz$ and the 
solid angle~$d\Omega$
in the background $\bar\dd(z)$, as they are set by observers.
The dimensionless fluctuation~$\Ddd$
in our observed data set is driven, not only by the 
fluctuation~$\dD$ in the observable~$\DDD$, but also by the fluctuation~$\de_g$
in the observed galaxy number density. In case there is no galaxy ($\de_g=-1$)
at a given position and hence no observation at all, 
the perturbation part is $\Ddd=-1$, implying
no observed data~$\dd=0$, consistent with our arguments. As stated in the
example of the luminosity distance observable, the continuous 
field~$\dd(\xvec)$ as our theoretical modeling of the observed data set
should not be considered as the luminosity distance itself at a given position,
but as a useful description of the observed data set in the limit~$N\RA\infty$,
accounting for the bias due to the host galaxy clustering.

When our cosmological observable is simply the galaxy number 
density~$n$, we can directly use \EQR{eq:theory} for galaxy clustering
by replacing $\bar\DDD=1$ and setting $\dD=0$ in \EQRS{eq:bgddd}
and~(\ref{eq:perddd}), where no further weight by the galaxy number counts
is needed. We summarize our notation in Table~\ref{tab:cases}.

\subsection{Observables in a single redshift bin}
\label{ssec:obsonez}
It is often the case that the observed data set is confined 
to a single redshift bin $z_i\in[z,z+\De z]$, as one considers a sub-sample 
at the single redshift bin out of the full observations, 
spanning a wide range of redshift. In this case,
our theoretical descriptions of the data  can
be  obtained by replacing the infinitesimal redshift bin size~$dz$
with the finite (constant) width~$\De z$ as 
\beeq
\label{eq:singzD}
\bar N=4\pi\De z\dN~,\Dquad \bar \dd(z)=\bar \DDD(z){d\Om\over4\pi}
=:\hdd(z)d\Omega~,\Dquad
\dd(\xvec)=\bar\DDD(z){d\Omega\over4\pi}\left[1+\Ddd(\xvec)\right]~,
\eneq
where $\De z$ drops out in~$\bar\dd$ and~$\dd$ due to~$\De z$ in~$\bar N$.
The angular average 
\beeq
\label{eq:theavg}
\left\langle\dd\right\rangle_\Om(z):=\bar\DDD(z)\int{d\Omega\over4\pi}
\left[1+\Ddd(\xvec)\right]~,
\eneq
can be compared to the observed average~ $\AVE{\dd}^\up{obs}$ in 
\EQR{eq:dataavg}. Note that the total number~$N$ of  observed galaxies as
the denominator in \EQR{eq:dataavg}
is already accounted for in our theoretical description. 

At linear order
in perturbations, the angle average of~$\Ddd(\xvec)$ is often equated with
an ensemble average and is ignored as its perturbation vanishes 
at the linear order. However, the angle average
\beeq
\label{eq:monozero}
\Ddd_0(z):=\int {d\Omega\over4\pi}~\Ddd(z,\Vang)\neq0
\eneq
is non-vanishing even at the linear order in perturbations, unless it is
further averaged over all the possible observer positions in the Universe, or 
other light cones \cite{MIYOET19}, assuming the Ergodic hypothesis. 
Indeed, the angle average is
referred to as the monopole perturbation from the general multipole
expansion
\beeq
\label{eq:alm}
\Ddd(\xvec)=\sum_{lm}a_{lm}(z)Y_{lm}(\Vang)~,\Dquad \Ddd_0(z)={a_{00}
\over\sqrt{4\pi}}~.
\eneq
In the standard redshift-space distortion power spectrum analysis
(see, e.g., \cite{KAISE87,COFIWE94,YOSE13a}),
the observed galaxy fluctuation $\de_z(\xvec)$
is fully decomposed into the monopole,
the quadrupole, and the hexadecapole, none of which
are zero.\footnote{Note, however, that 
the multipoles in the redshift-space distortion
are to be understood as those of 
${\hat {\bf k}}$ with respect to a fixed observer direction ${\hat {\bf n}}$, 
while we decompose our variables as functions of $\hat{\bf n}$ into their multipoles, hence these multipoles mean something very different.}

The cosmic microwave background (CMB) anisotropies also 
correspond to the case of observables in a single redshift bin, 
in which the observed redshift is zero (no measurements in other redshifts,
in practice):
\beeq
\label{eq:cmbobs}
\bar\DDD(z):=\bar T~,\Dquad 
\Ddd(\xvec)\equiv \dD(\xvec)\equiv\Theta(\Vang):={\de T(\Vang)
\over \bar T}~.
\eneq
CMB anisotropies are an unbiased observable ($\de_g\equiv0$), i.e no
 weight~$\de_g$ 
from the galaxy number counts is involved in the description.
However, there exists a subtlety associated with the background
CMB temperature~$\bar T$ at $z=0$, which is an input cosmological parameter
of the model that cannot be exactly determined by observations. 
The observed CMB temperature $\AVE{T}^\up{obs}$ in observations \cite{FIRAS94}
is indeed the angular average in \EQRS{eq:theavg} and~\eqref{eq:dataavg} 
of the temperature measurements on the sky, and it is different from the
background temperature~$\bar T$ again due to the monopole 
contribution~$\Theta_0$ of the perturbation. The same difference exists in the luminosity
distance, for example, between the background luminosity distance~$\bar D_L(z)$
and the angle average~$\AVE{\dd_L(z)}_\Om$. The former is a mathematical function
of a homogeneous and isotropic universe, and the latter involves
the perturbations from various effects. This difference arises as we have
a single light cone in the Universe \cite{MIYOET19}. Note also that the larger the sphere over which we take an angle average the smaller its 'cosmic variance'. Hence cosmic variance is most substantial at small redshift.

In the CMB literature, the observed CMB temperature~$\AVE{T}^\up{obs}$ is always
used, so that it is more convenient to re-arrange the expression for the
CMB temperature on the sky:
\beeq
T(\Vang)=\bar T\left[1+\Theta(\Vang)\right]=\bar T\left(1+\Theta_0\right)
\times{1+\Theta(\Vang)\over1+\Theta_0}\equiv
\left\langle T\right\rangle^\up{obs}\left[1+\Theta^\up{obs}(\Vang)\right]~,
\eneq
which defines the observed temperature and its anisotropies
\beeq
\label{eq:Tobs}
\AVE{T}^\up{obs}:={1\over N}\sum^N_{i=1}T(\Vang_i)
=\int{d\Om\over4\pi}~T(\Vang)\equiv\bar T(1+\Theta_0)~,\Dquad
1+\Theta^\up{obs}(\Vang):={1+\Theta(\Vang)\over1+\Theta_0}~.
\eneq
Note that since the monopole~$\Theta_0$ is already 
absorbed in~$\AVE{T}^\up{obs}$, there is no monopole $\Theta^\up{obs}_0\equiv0$
in observations. In section~\ref{ssec:CMB} we further discuss subtleties
in~$\Theta(\Vang)$ associated with gauge choice and observer frame.
Table~\ref{tab:cases} summarizes our theoretical description of the observed
data set in a single redshift bin.

\subsection{Angular (projected) observables}
\label{ssec:angobs}
In weak lensing observations, the cosmological observables~$\DDD$ are primarily
composed of the lensing convergence~$\kappa$, shear~$\gamma_1$, $\gamma_2$, 
and rotation~$\omega$. 
The proper relativistic description of the lensing observables has been
developed  recently~\cite{YOGRET18,GRYO18}, demonstrating the existence of additional 
relativistic effects missing in the standard weak lensing formalism
and resolving the subtle issues associated with gauge choice and physical
rotation (see also \cite{BONVI08,BEBOVE10,SCJE12a,SCJE12b}).
Here,  we are not concerned with the details of~$\Ddd$, but we
note that the lensing observables vanish in the background due to the
symmetry. Therefore, the continuous field of the observable quantity~$\DDD$
in \EQR{eq:DDD} is in this case modeled as
\beeq
\DDD(\xvec):=\dD(\xvec)~,
\eneq
without a background part,
and our theoretical description of the observed data set is then described as
\beeq
\label{eq:NOBG}
\dd(\xvec)=\dD(\xvec){dN(\xvec)\over N}=:\bar\dd(z)\Ddd(\xvec)~,
\eneq
where we defined the ``background''
\beeq
\label{eq:LENBB}
\bar\dd(z)={1\over\bar N}\dN(z)dz d\Om
={\bar n(z)\rbar^2(z)\over\bar N H(z)(1+z)^3}dz d\Om~,
\eneq
and the perturbation~$\Ddd$ (but instead of $1+\Ddd$)
\beeq
\label{eq:per2d}
\Ddd(\xvec)={\dD(\xvec)\left[1+\de_g(\xvec)\right]\over1+\de N}~.
\eneq
The perturbation part $\Ddd=\dD+\OO(2)$ is devoid of any contribution
of the galaxy number density fluctuation~$\de_g$ at the linear order,
and $\bar\dd(z)$ is nothing
but the (normalized) redshift distribution of the observed galaxy number 
density in the background.

Since these observables like the lensing observables
are often dominated by measurement errors such as 
shape noise or  redshift measurement errors, they are summed over the
line-of-sight direction to enhance the signal-to-noise ratio. This procedure
is simply accommodated in our theoretical description as
the projected observed data set
\beeq
\label{eq:theory2d}
\dd^\up{pro}(\Vang):=\int_z\dd(\xvec)=
\int dz\left({1\over\bar N}\dN\right)d\Om~\Ddd(\xvec)~,
\eneq
where we use the superscript to represent the quantities that are projected 
along the line-of-sight direction. Similarly, the projected galaxy number
density is often used to probe galaxy clustering, when the redshift 
measurements are obtained by noisier photo-$z$ measurements 
or for radio galaxies where redshift information is very uncertain.
The projected number $dN^\up{pro}(\Vang)$ of the observed 
galaxies in a solid angle~$d\Omega$ is simply related to the number of
the observed galaxies in a volume~$dV$, but projected as
\beeq
dN^\up{pro}(\Vang)=\int_z dN(\xvec)=\int_zn(\xvec)dV
=d\Omega\int dz{dN\over dzd\Omega}=:n^\up{2D}(\Vang)d\Omega~,
\eneq
where we introduce the angular galaxy number density $n^\up{2D}$ and note that
the angular galaxy number density is dimensionless. Therefore, our theoretical
description of the angular galaxy number density  is
\beeq
n^\up{2D}(\Vang)=\int dz\dN\left[1+\de_g(\xvec)\right]~.
\eneq
In case that our cosmological observable is simply the projected galaxy number 
density~$n^\up{2D}$, we can directly use \EQR{eq:theory2d} for angular
galaxy clustering by and setting $\dD=1$ in \EQR{eq:per2d}.
Table~\ref{tab:cases} summarizes our theoretical description of the angular 
observables.

\begin{table*}
\caption{Theoretical descriptions of the cosmological observables}
\begin{tabular}{cccc}
\hline\hline
\parbox[10cm][0.8cm]{10cm}{}
observable $\dd=\bar\dd(z)(1+\Ddd)$ &  $\bar\dd(z)$ & $\Ddd(\xvec)$
& Equations\\
\hline\hline
\parbox[10cm][0.8cm]{10cm}{}
three-dimensional observables &
${\bar\DDD(z)\bar n(z)\rbar^2(z)\over
 \bar NH(z)(1+z)^3}dz d\Om$
&
${\dD(\xvec)+\de_g(\xvec)-\de N+\dD(\xvec)\de_g(\xvec)
\over1+\de N}$ 
& \eqref{eq:bgddd}, \eqref{eq:perddd}\\
\hline
\parbox[10cm][0.8cm]{10cm}{}
galaxy clustering ($\bar\DDD=1$, $\dD=0$) &
${\bar n(z)\rbar^2(z)\over H(z)(1+z)^3}dz d\Om$ & $\de_g(\xvec)$ 
& \eqref{eq:bgddd}, \eqref{eq:perddd}\\
\hline
\parbox[10cm][0.8cm]{10cm}{}
single redshift bin ($dz=\De z$) & ${\bar\DDD(z)\over4\pi}d\Om$ &
${\dD(\xvec)+\de_g(\xvec)-\de N+\dD(\xvec)\de_g(\xvec)
\over1+\de N}$ &\eqref{eq:singzD}, \eqref{eq:perddd}\\
\hline
\parbox[10cm][0.8cm]{10cm}{}
CMB $T(\Vang)=\bar T(1+\Theta)$ 
& $\bar T{d\Om\over4\pi}$ & $\Theta(\Vang)$ & \eqref{eq:cmbobs}\\
\hline
\parbox[10cm][0.8cm]{10cm}{}
lensing observable $\dd=\bar\dd(z)\Ddd$
&  ${\bar n(z)\rbar^2(z)\over \bar NH(z)(1+z)^3}dz
 d\Om$ & ${\dD(\xvec)\left[1+\de_g(\xvec)\right]\over1+\de N}$
&\eqref{eq:NOBG}$-$\eqref{eq:per2d}\\
\hline\hline
\end{tabular}
\label{tab:cases}
\end{table*}

\section{Likelihood of Cosmological Observables on the Light Cone}
\label{sec:like}
Given the redshift~$z$ and the angular position~$\Vang$ of galaxies
and the cosmological observables~$\dd$ 
derived from the other properties of galaxies
such as the luminosity distance, lensing shear, and so on, we construct
our favorite statistics to test the underlying cosmological 
models, and the tests of our cosmological models against the observations
are performed through the likelihood analysis: A given
set of cosmological parameters is used to predict the likelihood
of the measurements, and combined with the priors of the
cosmological parameters, 
the posterior is computed for each cosmological parameter set, and this
procedure is repeated until the best parameter set that maximizes the posterior
is found (see, e.g., \cite{TETAHE97,BOJAKN98}).

This likelihood analysis is generic and standard in literature. Here
we reformulate the likelihood analysis,
taking into consideration that the cosmological observables are obtained
on the light cone. This can be contrasted to the standard method
in literature, in which the analysis
is performed as though the survey volume would be
 a hypersurface of simultaneity. 
As long as the survey volume is small, the standard method
is a good approximation
to the real observations, but the systematic errors grow as redshift
depth and the sky coverage increase. Here we develop the likelihood analysis 
on the light cone without such limitations. In particular, we will compute
the {\it maximum information} contained on the light cone.

\subsection{Gaussian probability distribution}
\label{ssec:gaussian}
Though our formalism is generally applicable to the likelihood analysis
on the light cone, we make a series of assumptions to simplify our analytic
calculations:
From now on, we will exclusively deal with the linear perturbations,
ignoring any higher-order perturbations, and we further assume that the 
linear
perturbations have vanishing statistical
 mean and they are Gaussian-distributed on average.
Under these assumptions, the two-point correlation will contain all the
information, and any connected $N$-point correlations with $N>2$ vanish.

To construct the probability distribution given the observed data set 
$\dd^\up{obs}_i$, we first consider the theoretical descriptions of the 
data  at a point~$\xvec_i$  and compute
their ensemble average 
\beeq
\dd(\xvec_i)=\bar\dd(z_i)\left[1+\Ddd(\xvec_i)\right]~,\Dquad
\mu_i:=\AVE{\dd(\xvec_i)}=\bar\dd(z_i)~,
\eneq
where the ensemble average is performed over the hypersurface of fixed
observed redshift (see \cite{YODU17} for the subtle 
gauge issues associated with the ensemble average in a usual coordinate). 
We now define the deviation~$\DDd$ from the mean and its covariance~$\CCC$:
\beeq
\DDd(\xvec_i):=\dd(\xvec_i)-\mu_i=\bar\dd(z_i)\Ddd(\xvec_i)~,\Dquad
\CCC_{ij}:=\AVE{\DDd(\xvec_i)\DDd(\xvec_j)}=\bar\dd(z_i)\bar\dd(z_j)
~\xi_{ij}~,
\eneq
where we defined the dimensionless two-point correlation function
\beeq
\xi_{ij}:=\AVE{\Ddd(\xvec_i)\Ddd(\xvec_j)}~.
\eneq
Remember that the deviation~$\DDd(\xvec_i)$ of the observed data from the mean
includes the background quantity~$\bar\dd(z_i)$ and the 
perturbation~$\Ddd(\xvec_i)$. For a Gaussian distribution, 
the two-point correlation function~$\xi_{ij}$ contains all the information.
In real observations, the observed data set~$\dd_i^\up{obs}$ is a set of
numbers associated to the observed positions~$\xvec_i$. The mean~$\mu_i$
and the covariance~$\CCC_{ij}$ are  predictions of our chosen model
at the observed positions~$\xvec_i$.

Last, we need the inverse covariance~$\KKK_{ij}$
to construct the probability distribution, defined by
\beeq
\label{eq:dortho}
\sum_j^N\CCC_{ij}\KKK_{jk}=\bdv{I}_{ik}~,
\eneq
where $\bdv{I}$ is an identity matrix. 
The Gaussian probability distribution of the observed data set~$\dd^\up{obs}_i$
can now be written as
\beeq
\label{eq:gausspdf}
\pdf={1\over\left[(2\pi)^N\det\CCC\right]^{1/2}}
\exp\left[-\frac12\sum_{ij}^N\DDd_i\KKK_{ij}\DDd_j\right]~.
\eneq
It is the probability that we measure the observed data set 
given our theoretical predictions, and its logarithm is referred to as the 
likelihood $\mathcal{L}:=-\ln\pdf$.
With our cosmological model and its parameters, we predict
the mean $\mu_i$, the covariance matrix~$\CCC_{ij}$, and
its inverse~$\KKK_{ij}$. As described, the likelihood analysis proceeds as
follows: We first choose a set of cosmological parameters and predict~$\mu_i$,
$\CCC_{ij}$ and~$\KKK_{ij}$ to compute~$\pdf$ or~$\cal L$. We then explore
the cosmological parameter space to maximize~$\pdf$, given the observed
data set.

\subsection{Maximum cosmological information on the light cone}
\label{ssec:max}
The probability distribution is maximized, only if we choose the best
cosmological model and its parameters. Even after the maximum is found,
however, the cosmological information of the observed data 
set~$\dd_i^\up{obs}$ in a given survey
is {\it not infinite}, but {\it finite}. Furthermore, 
given the survey volume, more information can be  extracted, if we 
make more observations. Then the key question naturally arises: 
``{\it What is the maximum cosmological
information content on the light cone (up to some redshift)?}'' 
This question has not been properly addressed in literature.

To compute the maximum cosmological information contained in the light
cone volume, we make a series of idealized assumptions: 
The observed data
set is free of any systematic errors or measurement errors and is
obtained from an {\it infinite} number ($N=\infty$)
of galaxies within the survey boundary (no shot-noise contribution 
$n_g=\infty$). Of course, we need to know the
correct cosmological model and parameters.
Under these assumptions, the (discrete) observed data set~$\dd_i^\up{obs}$ 
becomes a continuous field~$\dd(\xvec)$ in \EQR{eq:theory}, and 
the probability distribution in \EQR{eq:gausspdf}
can be trivially generalized to this idealized
case by replacing the
discrete sum with an integral. In this limit, the
covariance~$\CCC_{ij}$ and the correlation function~$\xi_{ij}$ are also
naturally promoted to the continuous fields, and the inverse covariance
is then subject to the continuous orthonormality condition, rather than
the discrete one in \EQR{eq:dortho}:
\beeq
\label{eq:inverse}
(\CCC~\KKK)(\xvec_1,\xvec_2)=
dz_1d\Omega_1\int dz'\int d\Omega'~\xi(\xvec_1,\xvec')\zeta(\xvec',\xvec_2)
=\de^D(z_1-z_2)\de^D(\Omega_1-\Omega_2)dz_1d\Omega_1~,
\eneq
where we replaced the Kronecker delta with the Dirac delta function, and
the indices $i,j,\cdots$ are also replaced with continuous field variables.
Note that both $\CCC$ and~$\KKK$ are infinite-dimensional matrices.
We also defined the ``inverse'' correlation function~$\zeta(\xvec_1,\xvec_2)$ 
with $\hdd$ (not with $\bar\dd$) as
\beeq
\KKK_{12}:=
\KKK(\xvec_1,\xvec_2):=\left[\hdd(z_1)\hdd(z_2)\right]^{-1}\zeta(\xvec_1,
\xvec_2)~,\Dquad \zeta_{12}:=\zeta(\xvec_1,\xvec_2)~,
\eneq
where we defined a short-hand notation for the infinite dimensional matrices.
From now on, we will work with the continuous fields to compute the
maximum cosmological information contents on the light cone.

\subsection{Decompositions and inverse covariance matrix}
\label{ssec:inverse}
The observation on the light cone is made in terms of the observed
redshift and  angular position, and the radial information from the
redshift represents not only the radial coordinate, but also the time
along the past light cone. This rather trivial
observation demands that we treat these
two fundamental observables {\it differently}. 
While statistical isotropy implies that harmonic modes in angular direction are statistically independent, radial modes are not,
 even if we assume statistical homogeneity, since on the light cone they mix spatial and time directions.

First, we compute the inverse covariance in a single redshift bin in terms of
spherical harmonics decomposition of the angular position, 
where the radial information is fixed or integrated
out. Second, we compute the inverse covariance utilizing the full 3D
information in terms of spherical Fourier decomposition.

\subsubsection{Single redshift bin: Angular decomposition}
\label{ssec:sininv}
As described in section~\ref{ssec:obsonez}, we may be interested in 
the observables
in a single redshift bin such as the type-Ia supernovae at a given redshift
and the CMB temperature anisotropies today. These observables are essentially
a function of angular position alone, and the angular decomposition based on
spherical harmonics provides a useful description.

In a single redshift bin, the observed data set is modeled in 
section~\ref{ssec:obsonez} and summarized in Table~\ref{tab:cases} as
\beeq
\label{eq:mu2d}
\dd(\xvec)=\bar\DDD(z){d\Om\over4\pi}\left[1+\Ddd(\xvec)\right]~,
\Dquad \mu=\bar\DDD(z){d\Om\over4\pi}~,\Dquad \hdd(z)={\bar\DDD(z)\over4\pi
\De z}~.
\eneq
The deviation from the mean, the covariance matrix, and the inverse
covariance matrix are then
\beeq
\label{eq:cov2d}
\DDd(\xvec)=\bar\DDD(z){d\Omega\over4\pi}~\Ddd(\xvec)~,\Dquad
\CCC_{12}=\bar\DDD^2(z){d\Omega_1\over4\pi}{d\Omega_2\over4\pi}
~\xi_{12}~,\Dquad
\KKK_{12}=\left({4\pi\De z\over\bar\DDD(z)}\right)^2\zeta_{12}~,
\eneq
where $\xvec_1=(z,\Vang_1)$ and $\xvec_2=(z,\Vang_2)$.
The orthonormality relation in \EQR{eq:inverse} can be integrated over 
the redshift width to derive the orthonormality relation in a single redshift
bin
\beeq
\label{eq:inverseang}
\int_z\int_{z'}(\CCC\KKK)_{12}=(\De z)^2
d\Omega_1\int d\Omega'~\xi(\xvec_1,\xvec')\zeta(\xvec',\xvec_2)
=\de^D(\Omega_1-\Omega_2)d\Omega_1~.
\eneq
To make further progress, we decompose the angular position~$\Vang$ of
the observable fluctuation in terms
of spherical harmonics~$Y_{lm}$ as
\beeq
\Ddd(\xvec):=\sum_{lm}a_{lm}(z)Y_{lm}(\Vang)~,\Dquad
a_{lm}(z)\equiv\int d\Omega~Y_{lm}^*(\Vang)\Ddd(\xvec)~,
\eneq
where $a_{lm}$ is the angular coefficient that depends on the redshift. 
Using statistical isotropy, the angular power spectrum is 
\beeq
\AVE{a_{lm}a^*_{l'm'}}=\de_{ll'}\de_{mm'}C_l~,
\eneq
and the two-point correlation function can be angular decomposed as
\beeq
\label{eq:xicl}
\xi_{12}=\AVE{\Ddd(\xvec_1)\Ddd(\xvec_2)}
=\sum_{lm}C_l Y_{lm}(\Vang_1)Y_{lm}^*(\Vang_2)
=\sum_l{2l+1\over4\pi}C_l P_l(\ga_{12})~,\Dquad \ga_{12}:=\Vang_1\cdot\Vang_2~,
\eneq
where $P_l$ is the Legendre polynomial of degree~$l$.
The inverse correlation function will be angular decomposed in the exactly
same way, defining the ``inverse'' angular power spectrum~$\II_l$
\beeq
\label{eq:zeta2d}
\zeta_{12}:=\sum_{lm}\II_lY_{lm}(\Vang_1)Y_{lm}^*(\Vang_2)=
\sum_l{2l+1\over4\pi}\II_lP_l(\ga_{12})~.
\eneq
By using the orthonormality condition in \EQR{eq:inverseang},
the inverse angular power spectrum can be readily obtained as
\beeq
\II_l={1\over C_l(\De z)^2}~,
\eneq
where we used 
\beeq
\sum_{lm}Y_{lm}(\Vang_1)Y_{lm}^*(\Vang_2)=\de^D(\Omega_1-\Omega_2)=
{1\over\sin\ttt_1}\de^D(\ttt_1-\ttt_2)\de^D(\pp_2-\pp_2)~.
\eneq
Indeed, the inverse angular power spectrum~$\II_l$ is the inverse of the 
angular power spectrum~$C_l$, and the inverse covariance is explicitly
\beeq
\label{eq:inverse2d}
\KKK_{12}={4\pi\over\bar\DDD(z)^2}\sum_l{2l+1\over C_l}
P_l(\ga_{12})~.
\eneq

\subsubsection{Light cone volume: Spherical Fourier decomposition}
\label{ssec:3dinv}
In addition to the angular position~$\Vang$, the cosmological observables
are marked by their radial position in terms of the observed redshift~$z$.
With this additional radial dimension, we perform the spherical Fourier 
decomposition. As opposed to the standard Fourier analysis in a rectangular 
coordinate, the spherical Fourier analysis utilizes the eigenfunctions 
of the Helmholtz equation in a spherical coordinate, which naturally
describes our fundamental observables~$(z,\Vang)$. 
The spherical Fourier analysis was
well developed \cite{BIQU91,FILAET95,HETA95} in galaxy clustering
and applied to the baryonic acoustic oscillation \cite{RARE12},
the three-dimensional weak lensing \cite{HEAVE03}, and
the relativistic effects \cite{YODE13}. Here we will adopt the formalism
and notation convention in \cite{YODE13}.

Given the three-dimensional positional
information of the cosmological observables
over the light cone volume, we apply the spherical Fourier analysis
to compute the spherical power spectrum, and the fluctuations in the
observable are now decomposed as
\beeq
\label{eq:3dFT}
\Ddd(\xvec):=\sum_{lm}\int_0^\infty dk\sqrt{2\over\pi}~kj_l(k\rbar)Y_{lm}
(\Vang)\sss_{lm}(k)~,
\eneq
where  $j_l(x)$ is the spherical Bessel function and
the spherical Fourier coefficients are
\beeq
\sss_{lm}(k)\equiv\int d\Omega\int d\rbar~\rbar^2\sqrt{2\over\pi}~kj_l(k\rbar)
Y_{lm}^*(\Vang)\Ddd(\xvec)~,
\eneq
where the radial integration is limited to the survey range.
Here we have assumed a fiducial cosmological model to relate the measured redshift to a radial distance $\rbar$. Hence, both $\rbar$ and $k$ are model dependent and not direct observables.
The amplitude square of the spherical Fourier coefficients~$\sss_{lm}(k)$ is
the spherical power spectrum
\beeq
\AVE{\sss_{lm}(k)\sss^*_{l'm'}(k')}=\de_{ll'}\de_{mm'}S_l(k,k')~,
\eneq
where we assume statistical isotropy.
The spherical power spectrum in the simplest case in Appendix~\ref{app:ob}
will be identical to the usual power spectrum $S_l(k,k')=\de^D(k-k')P(k)$.
Mind that the dimensions of the Fourier mode and its spherical
power spectrum 
are somewhat different from the usual Fourier analysis:
\beeq
[\sss_{lm}(k)]=L^2~,\Dquad [S_l(k,k')]=L^4~.
\eneq
In our spherical Fourier decomposition, we assumed that the sky coverage
is the full sky. In real surveys, the sky coverage is never the full sky, 
but our assumption is fine, because while the low angular multipoles 
in real surveys are inevitably correlated, the high angular multipoles 
are independent even with an incomplete sky coverage. 
The radial Fourier modes would be independent if the survey volume
is infinite and the fluctuations are time-translation invariant along
the line-of-sight direction. However, none of these two are true,
and we take into account that $S_l(k,k')$ is not diagonal since 
it mixes spatial and temporal information and our background universe 
is not time-translation invariant.

Given the spherical Fourier decomposition, we use the orthonormality condition
to derive the inverse covariance~$\KKK_{ij}$ or the inverse correlation 
function~$\zeta_{ij}$. We first compute the two-point correlation function 
in terms of the spherical power spectrum~$S_l(k,k')$:
\beeq
\label{eq:xi3d}
\xi_{12}=\AVE{\Ddd(\xvec_1)\Ddd(\xvec_2)}
=4\pi\sum_{lm}\int dk\int dk'~{kk'\over2\pi^2}S_l(k,k')
j_l(k\rbar_1)j_l(k'\rbar_2)Y_{lm}(\Vang_1)Y_{lm}^*(\Vang_2)~,
\eneq
and write the inverse correlation function in the similar way
\beeq
\label{eq:zetaFT}
\zeta_{12}:=4\pi\sum_{lm}\int dk\int dk'~{kk'\over2\pi^2}\IS_l(k,k')
j_l(k\rbar_1)j_l(k'\rbar_2)Y_{lm}(\Vang_1)Y_{lm}^*(\Vang_2)~,
\eneq
defining the ``inverse'' spherical power spectrum~$\IS_l(k,k')$. Due to the
rotational symmetry, both the correlation functions are only a function
of angular multipole~$l$,  but we keep the explicit dependence on~$\Vang_1$
and~$\Vang_2$ for later convenience.
Using the spherical Fourier decomposition of~$\xi_{12}$ and~$\zeta_{12}$
and integrating over~$d\Omega'$, 
the orthonormality condition in \EQR{eq:inverse} can be written as
\bear
\de^D(z_1-z_2)\de^D(\Om_1-\Om_2)&=&
(4\pi)^2\sum_{lm}Y_{lm}(\Vang_1)Y_{lm}^*(\Vang_2)
\int dk_1\int dk_2\int dk_1'\int dk_2'
~{k_1k_2\over2\pi^2}~{k_1'k_2'\over2\pi^2} \nnn
&&
\times
S_l(k_1,k_2)\IS_l(k_1',k_2')j_l(k_1\rbar_1)j_l(k_2'\rbar_2) \FF_l(k_2,k_1') ~,
\enar
where we defined the Fourier angular kernel
\beeq
\label{eq:FF}
\FF_l(k_1,k_2):=\int dz~j_l(k_1\rbar)j_l(k_2\rbar)~,
\eneq
and the integration range is again limited to the survey range.
Since the inverse spherical power spectrum is part of the integral equation,
we make a series of manipulations to pull out the inverse spherical
power spectrum from the integral by using the identities associated with
the Dirac delta function. First, we multiply the orthonormality equation
by $Y_{LM}(\Vang_2)$ and integrate  over~$d\Omega_2$ to obtain
\bear
&&(4\pi)^2\int dk_1\int dk_2\int dk_1'\int dk_2'
~{k_1k_2\over2\pi^2}~{k_1'k_2'\over2\pi^2} \nnn
&&\qquad\qquad
\times
S_l(k_1,k_2)\IS_l(k_1',k_2')j_l(k_1\rbar_1)j_l(k_2'\rbar_2) \FF_l(k_2,k_1') 
=\de^D(z_1-z_2)~,
\enar
where we re-labeled the angular index~$L$ by~$l$. It is noted that the
orthonormality condition after the angular integration becomes independent
of angular multipoles~$l$, or identical for all angular multipoles.
 To further simplify the condition, we use the mathematical
identity of spherical Bessel functions (independent of the survey
depth)
\beeq
\int_0^\infty d\rbar ~\rbar^2j_l(k_1\rbar)j_l(k_2\rbar)={\pi\over2k_1k_2}
\de^D(k_1-k_2)~,
\eneq
and integrate over $d\rbar_2$ after multiplying by $\rbar^2_2j_l(k_A\rbar_2)$
to derive
\beeq
4\pi\int dk_1\int dk_2\int dk_1'
~{k_1k_2\over2\pi^2}~{k_1'\over k_A} 
S_l(k_1,k_2)\IS_l(k_1',k_A)j_l(k_1\rbar_i) \FF_l(k_2,k_1') 
=\left({\rbar^2\over H}\right)_{z_1}j_l(k_A\rbar_1)~.
\eneq
Finally, we use the identity of the spherical Bessel function
 one more time by multiplying by
$j_l(k_B\rbar_1)$ and integrating over~$dz_1$,
and by re-arranging the equation, we arrive at the closed
 equation for the inverse spherical power spectrum
\beeq
\left({2\over\pi}\right)^2\int dk_1'~k_1'k_B\IS_l(k_1',k_A)\times
\int dk_1\int dk_2~k_1k_2S_l(k_1,k_2)
\FF_l(k_2,k_1')\FF_l(k_B,k_1)=\de^D(k_A-k_B)~.
\eneq
This integral equation is multiplications of infinite-dimensional matrices,
resulting in the identity matrix:
\beeq
\label{eq:masterIS}
\bdv{\tilde S}_l\left(\bdv{F}_l\bdv{S}_l\bdv{F}_l\right)=\bdv{I}~,
\Dquad
\bdv{\tilde S}_l=\left(\bdv{F}_l\bdv{S}_l\bdv{F}_l\right)^{-1}~,
\eneq
where we used the boldface letters for the matrices
\beeq
(\bdv{\tilde S}_l)_{12}=\IS_l(k_1,k_2)k_2~,\qquad
\left(\bdv{S}_l\right)_{12}=S_l(k_1,k_2)k_2~,\qquad
\left(\bdv{F}_l\right)_{12}={2\over\pi}~\FF_l(k_1,k_2)k_2~.
\eneq
The inverse spherical power spectrum should be obtained by inverting this 
matrix equation.
In section~\ref{ssec:gc}, we derive the inverse spherical power spectrum
$\IS_l(k,k')\propto P^{-1}(k)\de(k-k')$ in \EQR{eq:invSL} for a small survey volume,
where the flat-sky approximation and no time evolution can be adopted.
Therefore, the inverse covariance in the light cone volume is 
\beeq
\label{eq:inverse3d}
\KKK_{12}={4\pi\over\hdd(z_1)\hdd(z_2)}
\sum_l{2l+1\over4\pi}P_l(\ga_{12})\int dk\int dk'~{kk'\over2\pi^2}\IS_l(k,k')
j_l(k\rbar_1)j_l(k'\rbar_2)~.
\eneq

\subsubsection{Projected observables: Angular decomposition}
\label{ssec:invpro}
The observables in a single redshift bin in section~\ref{ssec:angobs}
were angular decomposed in section~\ref{ssec:sininv}, but they can also
be described in terms of spherical Fourier decomposition. The relation 
of the spherical power spectrum
to the angular power spectrum can be read off as
\beeq
\label{eq:single3d}
a_{lm}(z)=\int_0^\infty dk\sqrt{2\over\pi}kj_l(k\rbar_z)\sss_{lm}(k)~,
\Dquad
C_l(z)=
4\pi\int dk\int dk'~{kk'\over2\pi^2}S_l(k,k')j_l(k\rbar_z)j_l(k'\rbar_z)~.
\eneq
It is evident that the inverse covariance in \EQR{eq:inverse3d} over the
light cone volume is not the inverse covariance in \EQR{eq:inverse2d},
even when the survey volume is restricted to the sub-volume of
a single redshift bin, i.e., if more data beyond the single redshift bin
are available, we have to account for the radial correlation and more
information is available, even when
we restrict our analysis to a single redshift bin.

The projected observables such as the weak lensing observables
and the projected galaxy number density are somewhat different from 
what we derived for a single redshift bin
in section~\ref{ssec:sininv}, as the three-dimensional
spherical power spectrum is integrated along the line-of-sight direction.
Given the projected observable quantity~$\dd^\up{pro}(\Vang)$
in \EQR{eq:theory2d}, we construct the covariance
\beeq
\label{eq:covpro}
\CCC_{12}=d\Om_1 d\Om_2\int dz_1\left({1\over\bar N}\dN\right)_{z_1}
\int dz_2\left({1\over\bar N}\dN\right)_{z_2}
\AVE{\Ddd(\xvec_1)\Ddd(\xvec_2)}=:\xi_{12}d\Om_1 d\Om_2~,
\eneq
where the ensemble average in the integrand is the (three-dimensional)
two-point correlation function in \EQR{eq:xi3d} and we define the
(projected) angular two-point correlation~$\xi_{12}$, in comparison
to \EQR{eq:cov2d}. The angular two-point correlation function can be
further decomposed as in \EQR{eq:xicl}, and the angular power spectrum
of the projected observable~$\dd^\up{pro}$ is then
\beeq
\label{eq:clpro}
C_l=4\pi\int dk\int dk'~{kk'\over2\pi^2}S_l(k,k')
\left[\int dz_1\left({1\over\bar N}\dN\right)j_l(k\rbar_1)\right]\times
\left[\int dz_2\left({1\over\bar N}\dN\right)j_l(k'\rbar_2)\right]~.
\eneq
The angular power spectrum in \EQR{eq:single3d} is recovered, if the
redshift distribution is confined to a single redshift bin. Similarly,
the inverse covariance is defined in terms of the inverse correlation
function~$\zeta_{12}$
\beeq
\KKK_{12}:=\zeta_{12}=\sum_l{2l+1\over4\pi}\II_l P_l(\ga_{12})~,
\eneq
and the inverse angular power spectrum is
\beeq
\II_l={1\over C_l}~,
\eneq
in similarity to \EQR{eq:zeta2d}.

\section{Cosmological Information Contents on the Light Cone}
\label{sec:info}
As discussed in section~\ref{ssec:max}, we are interested in the maximum
possible cosmological information derivable from a given observable
measured in the light cone volume.
To compute these cosmological information contents, we employ the Fisher 
information formalism. The Fisher information technique has been well
developed for galaxy clustering and CMB analysis
(see, e.g., \cite{TETAHE97,BOJAKN98}). However,
since cosmological observables are mapped in terms of the redshift and
the angular position, the standard flat-sky or Euclidean descriptions
in the Fisher analysis
represent only the approximation to the real observations. Our Fisher matrix
analysis based on the redshift and the angular position provides the most
accurate description of the cosmological information contents in the light
cone volume, and further insights can be gained in connection to the
standard analysis in the limiting cases, where the survey volume is small
and narrow.

Moreover, our formalism in section~\ref{sec:observable} reveals 
that  cosmological observables come with  additional
fluctuations associated with the light propagation and the galaxy number 
counts. For instance, observations of supernova type Ia are 
used to measure the background luminosity distance~$\bar{\cal D}_L(z)$.
However, what we measure is indeed the full luminosity 
distance~$\dL(z)=\bar{\cal D}_L(z)(1+\ddL)$, including both the background
and the fluctuations,
as supernovae can go off only in some host galaxies (not in random places)
and the light propagation is thereafter affected by the fluctuations
between the source and the observer.
Consequently, even if we could measure an infinite number of supernovae
in a single redshift bin, the uncertainty in our luminosity distance 
measurements is limited by the cosmic variance, and the maximum cosmological
information is certainly {\it not} infinite, even in this idealized situation.
We will compute this maximum cosmological information contents on the light
cone.

First, we briefly review the Fisher information technique, and we then
proceed to compute the cosmological information contents in a single redshift 
bin and on the  light cone.

\subsection{Fisher information matrix}
\label{ssec:fisher}
The Fisher information formalism has been well developed and applied in
cosmology (see, e.g., \cite{TETAHE97,BOJAKN98}), and we provide a short
description to fix our notational convention. 

To compute the derivatives of the log likelihood ${\cal L}=-\ln\pdf$
around the maximum with
respect to the model parameters~$p_\mu$, we first construct the data matrix
out of the individual data~$\dd_i^\up{obs}$ and the model prediction~$\mu_i$
\beeq
\DM:=(\bm{\dd-\mu})(\bm{\dd-\mu})^\up{t}=\DDd_i\DDd_j~,
\eneq
where we used the boldface letter to indicate that the quantities are vectors and matrices. Choosing cosmological parameters such that $\bm\mu$ is
the best fit to the data $\bm{\dd}$, the ensemble average of first derivative of the data matrix vanishes,
\beeq
\AVE{{\pa\over\pa p_\mu}\DM}=0~,
\eneq
but the second derivative of the data matrix carries important information 
about the variation of the mean with respect to the model parameters
\beeq
\MMM_{\mu\nu}:=\AVE{{\pa^2\DM\over\pa p_\mu\pa p_\nu}}
={\pa\bm{\mu}\over\pa p_\mu}{\pa\bm{\mu}^\up{t}\over\pa p_\nu}
+{\pa\bm{\mu}\over\pa p_\nu}{\pa\bm{\mu}^\up{t}\over\pa p_\mu}~.
\eneq
Given the Gaussian probability distribution in \EQR{eq:gausspdf},
we compute the log-likelihood 
\beeq
-2\ln\pdf=\ln\det\CCC+\sum_{ij}\DDd_i\KKK_{ij}\DDd_j+N\ln 2\pi~,
\eneq
and the Fisher information matrix is then
\beeq
\label{eq:fisher}
F_{\mu\nu}:=\left\langle-{\pa^2\ln\pdf\over\pa p_\mu\pa p_\nu}\right\rangle
=\frac12\Tr\left[\left(\KKK{\pa\over\pa p_\mu}\CCC\right)
\left(\KKK{\pa\over\pa p_\nu}\CCC\right)+\KKK\MMM_{\mu\nu}\right]~.
\eneq
To leading order, the constant likelihood surface in parameter space
is determined by two distinct
contributions: the variations of the mean and the two-point correlation,
given the inverse covariance. The inverse of the Fisher matrix is
the optimistic forecast for the parameter estimation, representing the
cosmological information in the observed data set. We again generalize this
Fisher matrix formalism for the (discrete) observed data set~$\dd_i^\up{obs}$
to a continuous field~$\dd(\xvec)$.

\subsection{Single redshift bin and projected observables}
\label{ssec:infoz}
Cosmological observables in a single redshift bin or the projected
observables are decomposed in terms of spherical harmonics, and we have derived
the inverse covariance matrix in section~\ref{ssec:inverse}. We compute
the Fisher information matrix in~\EQR{eq:fisher}, using the angular
decomposition.

Given the cosmological observable~$\dd(\xvec)$ and its mean~$\mu$ in 
\EQR{eq:mu2d} in a single redshift bin, we first compute the variation
of the mean with respect to the model parameters
\beeq
\left(\MMM_{\mu\nu}\right)_{12}=2~{\pa\bar\DDD(z)\over\pa p_\mu}
{\pa\bar\DDD(z)\over\pa p_\nu}{d\Omega_1\over4\pi}{d\Omega_2\over4\pi}~,
\eneq
and its product with the inverse covariance is 
\beeq
\frac12\left(\KKK\MMM_{\mu\nu}\right)_{13}=\left(\De z\right)^2
{\pa\ln\bar\DDD(z)\over\pa p_\mu}
{\pa\ln\bar\DDD(z)\over\pa p_\nu}~d\Omega_3
\int d\Omega_2~\zeta_{12}~.
\eneq
Therefore, the first contribution to the Fisher information matrix can be
obtained by taking the trace and using the angular power spectrum
decomposition in \EQR{eq:zeta2d} as
\beeq
\label{eq:fisher2dA}
\frac12\Tr\bigg[\KKK\MMM_{\mu\nu}\bigg]\equiv\left(\De z\right)^2
{\pa\ln\bar\DDD(z)\over\pa p_\mu}
{\pa\ln\bar\DDD(z)\over\pa p_\nu}\int~d\Omega_1
\int~d\Omega_2~\zeta_{12}={4\pi\over C_0}\left(
{\pa\ln\bar\DDD(z)\over\pa p_\mu}\right)\left({\pa\ln\bar\DDD(z)\over\pa p_\nu}
\right)~,
\eneq
where $C_0$ is the monopole of the angular  power spectrum~$C_l$ and happens
to  be the cosmic variance of~$\bar\DDD(z)$ \cite{MIYOET19}. Here we see
that $C_0$ limits indeed the measurements of~$\bar\DDD(z)$ from the
Fisher matrix. We emphasize that the cosmic
variance of the background quantity in the measurements is {\it not}
appreciated in literature. The reason is that one often equates the angular
average with the ensemble average, such that the monopole contribution~$\Ddd_0$
is set zero, but as shown in \EQR{eq:monozero} this is {\it not} the case.
The equality holds, only when we perform the angular average over many
different observer positions, assuming the Ergodic hypothesis \cite{MIYOET19}.

The other contribution to the Fisher information matrix is the variation
of the covariance matrix with respect to the model parameters. The covariance
matrix in a single redshift bin is given in \EQR{eq:cov2d}, and its product
with the inverse covariance can be obtained as
\bear
\left(\KKK{\pa\over\pa p_\mu}\CCC\right)_{13} &\equiv&
{(\De z)^2\over\bar\DDD^2(z)}\int~d\Omega_2
~\zeta_{12}{\pa\over\pa p_\mu}\left[\bar\DDD^2(z)\xi_{23}
\right]d\Om_3 \nonumber \\
&=&\sum_{lm}Y_{lm}(\Vang_1)
{\pa\over\pa p_\mu}\bigg(\ln\left[\bar\DDD^2(z)C_l\right]\bigg)
Y^*_{lm}(\Vang_3)d\Om_3~,
\enar
where we used the angular power spectrum decomposition for~$\xi_{12}$ and
$\zeta_{12}$. By multiplying the same product with parameter~$p_\nu$
and taking the trace of the product, we derive the covariance contribution 
to the Fisher information matrix
\beeq
\label{eq:fisher2dB}
\frac12\Tr\left[\left(\KKK{\pa\over\pa p_\mu}\CCC\right)\left(\KKK
{\pa\over\pa p_\nu}\CCC\right)\right]=\sum_l{2l+1\over2}
{\pa\over\pa p_\mu}\bigg(\ln\bar\DDD^2(z)C_l\bigg)
{\pa\over\pa p_\nu}\bigg(\ln\bar\DDD^2(z)C_l\bigg)~.
\eneq
The covariance~$\CCC$ of the cosmological observable~$\dd(\xvec)$ contains 
the extra information about the observable, fully characterized by the 
two-point correlation function~$\xi(\xvec_1,\xvec_2)$ under the assumption
of Gaussianity. In a single redshift bin, this information is better 
represented by the angular power spectrum~$C_l$, and we recover the standard
expression, often phrased as follows: The angular power spectrum measurements
are limited as we can only sample $2l+1$ independent components of~$a_{lm}$.
This statement is true, but with one subtle caveat that we {\it cannot}
directly measure $\Ddd(\xvec)$ or its angular component $a_{lm}$ 
in \EQR{eq:alm}. What we can
measure is the full cosmological observable~$\dd(\xvec)$ that includes both the
background~$\bar\dd(z)$ and the perturbation~$\Ddd(\xvec)$, such that
our observed mean is limited by~$C_0$ and our observed power spectrum is 
limited by~$C_l$ but only through the combination $\bar\DDD^2(z) C_l$.

The full Fisher information matrix in a single redshift bin is then
\beeq
\label{eq:fisher2d}
F_{\mu\nu}={4\pi\over C_0}\left(
{\pa\ln\bar\DDD(z)\over\pa p_\mu}\right)\left({\pa\ln\bar\DDD(z)\over\pa p_\nu}
\right)+\sum_l{2l+1\over2}
{\pa\over\pa p_\mu}\bigg(\ln\bar\DDD^2(z)C_l\bigg)
{\pa\over\pa p_\nu}\bigg(\ln\bar\DDD^2(z)C_l\bigg)~.
\eneq
It represents the maximum information contained in the observed data 
set~$\dd^\up{obs}_i$ in a single redshift bin under the assumption that
the underlying fluctuations are Gaussian distributed. This result has been
well-known in literature without the background part~$\bar\DDD(z)$
(see, however, \cite{TETAHE97}).
For the projected observables in section~\ref{ssec:angobs}, where the 
observables vanish in the background, the Fisher information matrix 
contains only the covariance part
\beeq
\label{eq:fisher2dN}
F_{\mu\nu}=\sum_l{2l+1\over2}\left(
{\pa\over\pa p_\mu}\ln C_l\right)\left({\pa\over\pa p_\nu}\ln C_l\right)~,
\eneq
where the angular power spectrum~$C_l$
of the projected observables is given in \EQR{eq:clpro}.

\subsection{Observations of the three-dimensional light cone volume}
\label{ssec:infovol}
With the radial information given by the observed redshift,
 cosmological observables in a light cone volume
are decomposed in terms of spherical harmonics and spherical Bessel functions,
and we have derived
the inverse covariance matrix in section~\ref{ssec:3dinv}. Here we compute
the Fisher information matrix in~\EQR{eq:fisher}, using the spherical
Fourier decomposition, and this provides the most accurate description
of the cosmological information contents on a light cone.

The calculations on the light cone proceed in a similar way to those
in a single redshift bin in section~\ref{ssec:infoz}. We compute the matrix 
multiplications for the
Fisher information matrix and perform the spherical Fourier decomposition,
instead of the angular decomposition used in section~\ref{ssec:infoz}.
The variation of the mean with respect to the model parameters is
\beeq
\left(\MMM_{\mu\nu}\right)_{12}=\left[{\pa\over\pa p_\mu}\hdd(z_1)
{\pa\over\pa p_\nu}\hdd(z_2)+{\pa\over\pa p_\nu}\hdd(z_1)
{\pa\over\pa p_\mu}\hdd(z_2)\right]dz_1 d\Om_1 dz_2 d\Om_2~,
\eneq
and its product with the inverse covariance is
\beeq
\frac12\left(\KKK\MMM_{\mu\nu}\right)_{13}={dz_3d\Om_3\over2\hdd(z_1)}
\int dz_2\int d\Om_2~\zeta_{12}\left[{\pa\over\pa p_\mu}\ln\hdd(z_2)
{\pa\over\pa p_\nu}\hdd(z_3)+{\pa\over\pa p_\nu}\ln\hdd(z_2)
{\pa\over\pa p_\mu}\hdd(z_3)\right]~.
\eneq
Taking the trace of the product and using the spherical Fourier decomposition
in \EQR{eq:zetaFT}, we derive the first contribution to the Fisher information
matrix in the three-dimensional light cone volume
\bear
\label{eq:fisher3dA}
\frac12\Tr\bigg[\KKK\MMM_{\mu\nu}\bigg]&\equiv&
\int dz_1\int d\Om_1 \int dz_2 \int d\Om_2
~\frac12\zeta_{12}\left[\left({\pa\over\pa p_\mu}\ln\hdd(z_1)\right)\left(
{\pa\over\pa p_\nu}\ln\hdd(z_2)\right)
+\left(1\leftrightarrow 2\right)\right] \nnn
&=&(4\pi)^2\int dk\int dk'~{kk'\over2\pi^2}\IS_0(k,k')\GD_\mu(k)\GD_\nu(k')~,
\enar
where we used the spherical Fourier decomposition and 
defined the Fourier kernel
\beeq
\label{eq:GD}
\GD_\mu(k):=\int dz~j_0(k\rbar){\pa\over\pa p_\mu}\ln\hdd(z)~.
\eneq
Note that $\bar r$ depends on the cosmological model.

Compared to \EQR{eq:fisher2dA}, the contribution of the variation in the mean
to the Fisher matrix information takes the similar structure: the variation
of the mean value with respect to the model parameters is
limited by the inverse monopole power spectrum~$\IS_0(k,k')$. 
The Fisher information matrix shows
that there exists the cosmic variance in the background quantity~$\hdd(z)$,
again set by the monopole ($l=0$), but integrated over different Fourier
modes in the light cone volume.

We then move to compute the variation of the covariance matrix with respect
to the model parameters. First, we compute the product of the inverse
covariance and the derivative of the covariance matrix
\bear
\left(\KKK{\pa\over\pa p_\mu}\CCC\right)_{13}&=&{dz_3 d\Om_3\over\hdd(z_1)}
\int dz_2\int d\Om_2~
\left[{\zeta_{12}\over \hdd(z_2)}\right]
{\pa\over \pa p_\mu}\left[\hdd(z_2)\hdd(z_3)\xi_{23}\right] \\
&=&
(4\pi)^2\sum_{lm}{Y_{lm}(\Vang_1)Y_{lm}^*(\Vang_3)\over\hdd(z_1)}dz_3 d\Om_3
\int dk_1\int dk_2{k_1k_2\over2\pi^2}\IS_l(k_1,k_2)j_l(k_1\rbar_1)\nnn
&&\qquad\times\int dz_2{j_l(k_2\rbar_2)\over\hdd(z_2)}
\int dk_1'\int dk_2'{k_1'k_2'\over2\pi^2}{\pa\over\pa p_\mu}
\left[\hdd(z_2)\hdd(z_3)j_l(k_1'\rbar_2)j_l(k_2'\rbar_3)S_l(k_1',k_2')
\right]~,\nn
\enar
where we integrated over~$d\Om_2$. By repeating the calculation with the
parameter~$p_\nu$ and taking the trace of the product, we obtain the other
contribution to the Fisher information matrix
\bear
\label{eq:inter}
&&\frac12\Tr\left[\left(\KKK{\pa\over\pa p_\mu}\CCC\right)\left(\KKK
{\pa\over\pa p_\nu}\CCC\right)\right]=(4\pi)^4\sum_l{2l+1\over2}
\int dz_1\int dz_2\int dz_3\int dz_4 \\
&&\qquad\qquad\times
\int dk_1\int dk_2\int dk_3\int dk_4\int dk_1'\int dk_2'\int dk_3'\int dk_4'~
{k_1k_2\over2\pi^2}{k_3k_4\over2\pi^2}{k_1'k_2'\over2\pi^2}
{k_3'k_4'\over2\pi^2}\nnn
&&\qquad\qquad\times
{\IS_l(k_1,k_2)\over\hdd(z_1)\hdd(z_2)}~j_l(k_1\rbar_1)j_l(k_2\rbar_2)
{\pa\over\pa p_\mu}\left[\hdd(z_2)\hdd(z_3)j_l(k_1'\rbar_2)j_l(k_2'\rbar_3)
S_l(k_1',k_2')\right]\nnn
&&\qquad\qquad\times
{\IS_l(k_3,k_4)\over\hdd(z_3)\hdd(z_4)}~j_l(k_3\rbar_3)j_l(k_4\rbar_4)
{\pa\over\pa p_\nu}\left[\hdd(z_4)\hdd(z_1)j_l(k_3'\rbar_4)j_l(k_4'\rbar_1)
S_l(k_3',k_4')\right]~.\nn
\enar
To simplify the expression, we define a series of Fourier kernels,
in addition to the Fourier angular kernel~$\FF_l(k_1,k_2)$ in \EQR{eq:FF}:
\bear
\FF_l(k_1,k_2)&:=&\int dz~j_l(k_1\rbar)j_l(k_2\rbar)=:\FF_l^{12}~,\\
\HK_{l,\mu}(k_1,k_2)&:=&\int dz~j_l(k_1\rbar)j_l(k_2\rbar){\pa\over\pa p_\mu}
\ln\hdd(z)=:\HK_{l,\mu}^{12}~,\\
\NK_{l,\mu}(k_1,k_2)&:=&\int dz~j_l(k_1\rbar){\pa\over\pa p_\mu}j_l(k_2\rbar)
=:\NK_{l,\mu}^{12}~,
\enar
where we used the super-scripts to simplify the arguments and
the kernels~$\FF_l(k_1,k_2)$ and~$\HK_{l,i}(k_1,k_2)$
are symmetric in their arguments, but~$\NK_{l,\mu}(k_1,k_2)$ is not.
Expanding the derivatives in \EQR{eq:inter} and integrating over the
redshift, we derive
\bear
\label{eq:fisher3dB}
&&\frac12\Tr\left[\left(\KKK{\pa\over\pa p_\mu}\CCC\right)\left(\KKK
{\pa\over\pa p_\nu}\CCC\right)\right]=\left({2\over\pi}\right)^4
\sum_l{2l+1\over2}
\left(\prod_{i=1}^4\int dk_i~k_i\right)\left(\prod_{j=1}^4\int dk_j'~k_j'
\right)\IS_l(k_1,k_2)\IS_l(k_3,k_4)
\nnn
&&\qquad\times
\left[\FF_l^{21'}\bigg(\HK_{l,\mu}^{32'}+\NK_{l,\mu}^{32'}\bigg)S_l(k_1',k_2')+
\FF_l^{32'}\bigg(\HK_{l,\mu}^{21'}+\NK_{l,\mu}^{21'}\bigg)S_l(k_1',k_2')
+\FF_l^{21'}\FF_l^{32'}{\pa\over\pa p_\mu}S_l(k_1',k_2')\right]\nnn
&&\qquad\times
\left[\FF_l^{43'}\bigg(\HK_{l,\nu}^{14'}+\NK_{l,\nu}^{14'}\bigg)S_l(k_3',k_4')+
\FF_l^{14'}\bigg(\HK_{l,\nu}^{43'}+\NK_{l,\nu}^{43'}\bigg)S_l(k_3',k_4')
+\FF_l^{43'}\FF_l^{14'}{\pa\over\pa p_\nu}S_l(k_3',k_4')\right]~.~~~~~~~~~~~
\enar
Compared to \EQR{eq:fisher2dB}, the contribution of the covariance to the
Fisher information matrix is substantially more complicated, as it involves
the three-dimensional spherical Fourier decomposition. However, the structure
is similar in a sense that the three-dimensional fluctuations are correlated
not only in angular directions, but also in radial direction, so that
the measurements are limited by the cosmic variance given by the inverse
spherical power spectrum~$\IS_l(k,k')$. Furthermore, it clearly shows that
the cosmological information is contained not only in the spherical power 
spectrum~$S_l(k,k')$, but also in the angular diameter distance~$\rbar$
and the background mean~$\hdd(z)$ through the Fourier kernels~$\HK_{l,\mu}$
and~$\NK_{l,\mu}$.

Adding the two contributions, 
the full Fisher information matrix on the  light cone can be written as
\bear
\label{eq:fisher3d}
F_{\mu\nu}
&=&(4\pi)^2\int dk\int dk'~{kk'\over2\pi^2}\IS_0(k,k')\GD_\mu(k)\GD_\nu(k') \\
&+&\left({2\over\pi}\right)^4\sum_l{2l+1\over2}
\left(\prod_{i=1}^4\int dk_i~k_i\right)\left(\prod_{j=1}^4\int dk_j'~k_j'
\right)\IS_l(k_1,k_2)\IS_l(k_3,k_4)\nnn
&&\times
\left[\FF_l^{21'}\bigg(\HK_{l,\mu}^{32'}+\NK_{l,\mu}^{32'}\bigg)S_l(k_1',k_2')+
\FF_l^{32'}\bigg(\HK_{l,\mu}^{21'}+\NK_{l,\mu}^{21'}\bigg)S_l(k_1',k_2')
+\FF_l^{21'}\FF_l^{32'}{\pa\over\pa p_\mu}S_l(k_1',k_2')\right]\nnn
&&\times
\left[\FF_l^{43'}\bigg(\HK_{l,\nu}^{14'}+\NK_{l,\nu}^{14'}\bigg)S_l(k_3',k_4')+
\FF_l^{14'}\bigg(\HK_{l,\nu}^{43'}+\NK_{l,\nu}^{43'}\bigg)S_l(k_3',k_4')
+\FF_l^{43'}\FF_l^{14'}{\pa\over\pa p_\nu}S_l(k_3',k_4')\right]~.\nn
\enar
This Fisher matrix represents the maximum cosmological information
derivable from the observable in a light cone volume, again under the 
assumption that the underlying fluctuations are Gaussian distributed
(see section~\ref{ssec:max}). While the formal equation~\eqref{eq:fisher}
for the Fisher information matrix was well known, this 
equation~\eqref{eq:fisher3d} in a three-dimensional light-cone volume
is derived for the first time in this work.

Neglecting the terms $\HK_{l,\nu}$ and $\NK_{l,\nu}$ and using Limber approximation for the integrals over redshift (i.e. neglecting correlations at different redshifts) we obtain the well known results in 3D Fourier space. We shall show this in more details when treating examples in Section~\ref{sec:app}.

Before we terminate this section, we comment on the model-dependence of the
power spectrum analysis (both the traditional and the spherical). The analysis
involves a conversion of the observed galaxy position $\xvec=(z,\Vang)$
into the comoving distance~$\rbar$, in which we need a prior cosmological
model. In principle, this poses {\it no} problem, as the data processing
in this case is part of the likelihood analysis, in which the raw observed
data is re-processed for each set of cosmological parameters and
compared to the theoretical predictions (see, e.g., \cite{TETAHE97,TEEIET06}).
In practice, however, it is computationally more expensive to re-process the
raw data for each likelihood ladder and hence a simple approximation
is typically adopted (see, e.g., \cite{SEEI03,EIZEET05}).
In Appendix~\ref{app:sp}, we present an alternative to the power spectrum
analysis based on~$\rbar$ by using the observed redshift itself
as a dimensionless radial coordinate, and in this way the observed raw data
can be processed only once in a model independent way.

\section{Cosmological Applications}
\label{sec:app}
We apply our formalism to five different cosmological observables
to compute the maximum cosmological information contents derivable from
such observations in the idealized case described in section~\ref{ssec:max}.
Here we make a series of approximations and derive rough analytical 
estimates of the maximum cosmological information contents. A
 detailed analysis will require more extensive numerical investigations
beyond our current scope.

\subsection{Luminosity distance measurements 
in a single redshift bin with infinite number of supernovae}
\label{ssec:lumi}
Consider observations of luminous type-Ia supernovae in a single redshift bin
in a future survey, where a large number~$N$ of supernovae will be measured.
Each measurement provides an estimate of the luminosity 
distance~$\bar D_L(z)$, 
but the estimate is dominated by the intrinsic scatter due to the variation
of the absolute luminosity. The measurement uncertainty on the luminosity
distance can be beaten down by $N$-independent measurements of
supernovae at the same redshift, such that we expect to measure the
background luminosity distance~$\bar D_L(z)$ precisely, provided that
the number~$N$ of supernovae is sufficiently large ($N\RA\infty$).

However, this standard picture is {\it incorrect}: What we measure
is not the background luminosity distance~$\bar D_L(z)$, but the number 
weighted luminosity distance~$\dd(\xvec)$, where our
observable~$\DDD(\xvec)$ in \EQR{eq:DDD} corresponds to the luminosity 
distance including the perturbation and the
observed data~$\dd_i^\up{obs}$ in \EQR{eq:dataset}
is the set of luminosity distances 
weighted by the host galaxy number count, as described 
by~$\dd(\xvec)$ in \EQRS{eq:ddNN} and~(\ref{eq:theory}).
We emphasize again that the supernova measurements include not only the
background luminosity distance, but also the perturbations in \EQR{eq:perddd},
which represent the fluctuations in the luminosity distance and the host 
galaxy. Because of this, even in this idealized situation,
where we can beat down the intrinsic
scatter completely with infinite number of supernova observations,
there exists a cosmic variance limit to the
luminosity distance measurements.
Therefore, it is important to compute this cosmic variance limit
(or the maximum cosmological information contents) derivable
from  supernova observations in a single redshift bin.\footnote{In
\cite{BEGAET13,BEDUET14}, similar arguments are presented, regarding
the cosmic variance limit. They computed the cosmic variance on the angle 
average $\AVE{\dd_L}_\Om(z)$ of the luminosity distances, while ignoring
the linear-order monopole contribution and the host galaxy fluctuation,
but focusing on the second-order relativistic
contributions. So, the cosmic variance
obtained in \cite{BEGAET13,BEDUET14} is smaller than our estimate 
in this section.}

This maximum cosmological information content is derived in \EQR{eq:fisher2d}
under the assumption that the underlying fluctuations are Gaussian and linear.
Focusing on the background part, we notice that the variance on the background
measurements is the monopole power spectrum~$C_0$: 
\beeq
F_{\mu\nu}\propto {4\pi\over C_0}\left(
{\pa\ln\bar\DDD(z)\over\pa p_\mu}\right)\left({\pa\ln\bar\DDD(z)\over\pa p_\nu}
\right)~,
\eneq
as in \EQR{eq:fisher2dA}. Observers derive the
best estimate of the background luminosity distance by averaging over
the sky coverage in \EQR{eq:dataavg}, and this estimate is represented by
$\AVE{\dd}_\Om(z)$ in
\EQR{eq:theavg}, which includes the background luminosity distance and
the monopole perturbation~$\Ddd_0(z)$ in \EQR{eq:monozero}. The monopole
is correlated
\beeq
\AVE{\Ddd_0\Ddd_0}\equiv {\AVE{|a_{00}|^2}\over4\pi}={C_0\over4\pi}~,
\eneq
and since we only have access to one light cone, the monopole power spectrum
sets the cosmic variance limit to the background measurements.

Once we account for the fact that our luminosity distance measurements
include perturbations, it is inevitable that the measurements are limited by
cosmic variance. The standard way of estimating  cosmic variance is
to compute 
\beeq
\sigma^2_\up{std}:=\AVE{\Ddd(\xvec)\Ddd(\xvec)}\equiv\xi(0)~,
\eneq
which we refer to as the standard variance. However, note that the fluctuations
of the host galaxies are often ignored in literature. 
Using the angular decomposition in \EQR{eq:xicl}, we derive 
\beeq
\sigma^2_\up{std}=\xi(0)=\sum_l{2l+1\over4\pi}C_l \geq{C_0\over4\pi}~,
\eneq
where we used $P_l(1)=1$. Given the correct formula for~$\Ddd$,
the standard variance is indeed larger than the
real cosmic variance limit $C_0/4\pi$, because $\sigma^2_\up{std}$ is
the variance of the luminosity distance fluctuations at each spatial point, 
while
the variance we need for the background estimate is the variance
on the angle-averaged fluctuations. We can gain further insight by computing
the cosmic variance in configuration space
\beeq
\label{eq:xiint}
\AVE{\Ddd_0\Ddd_0}:=\xi_0=
\int{d\Om_{12}\over4\pi}~\xi(\Vang_{12})=\frac12\int_0^\pi d\ttt~\sin\ttt~
\xi\left(r=2\rbar_z\sin{\ttt\over2};z\right)\leq\xi(0)~.
\eneq
The monopole correlation~$\xi_0(z)$ is literally the angle average of the
full three-dimensional correlation function~$\xi(r;z)$ on the light cone, and
we note that the equality holds only at the tip of the light cone ($z=0$).

Without detailed calculations, we can obtain a rough estimate of the cosmic 
variance limit. In \EQR{eq:perddd}, the perturbation~$\Ddd(\xvec)$ in the
observed data set is composed of the fluctuation~$\dD$
in the luminosity distance and the fluctuation~$\de_g$ 
in the host galaxy clustering.
The fluctuation in the luminosity distance was computed, properly accounting
for the relativistic contributions 
(see, e.g., \cite{BODUGA06,BODUKU06,YOSC16,BIYO16,SCYO17,BIYO17}), 
where it was shown that 
the velocity contributions are larger than the gravitational potential 
contribution. So, it is clear that the dominant contribution
to the supernova observations comes from the fluctuation in the host galaxy
clustering, which is in proportion to the matter density fluctuation.
However, this density contribution to the monopole~$\Ddd_0$ is cancelled
by~$\de N$, so that the leading contribution to the cosmic variance is
the velocity. Considering a simple estimate~$\sigma^2_v\simeq10^{-5}$---$10^{-4}$
at low redshift, a percent level cosmic variance is expected, while at
high redshift the lensing contribution is larger than the velocity 
contribution (see also \cite{HUGR06,BEGAET13,BEDUET14},
where similar results have been obtained).

Supernova observations, however, are not limited to a single redshift bin,
but in general cover a range of redshifts. Furthermore, we are somewhat less interested in
the background luminosity distances~$\bar D_L(z)$ at each redshift, but
more in deriving cosmological parameter constraints from the
supernova measurements over the redshift ranges.
One must therefore consider the full $\dd_L(z)$ function and the full 
light-cone Fisher analysis of section~\ref{ssec:infovol}.
As emphasized, even with an infinite number of supernova
observations, we cannot perfectly measure the background luminosity distances
over a range of redshifts, and there {\it exist} a maximum cosmological
information content set by the cosmic variance limit. To put it differently,
there exists a minimum error for cosmological parameter estimation
 from supernova observations
as we have only one light cone at our disposition (up to a maximum redshift).
This question will be investigated in detail in future work \cite{wedo}.

\subsection{Cosmic microwave background anisotropies: Do we know the
background CMB temperature?}
\label{ssec:CMB}
In observations of  cosmic microwave background anisotropies, we measure
the CMB temperature~$T(\Vang)$ as a function of angular position~$\Vang$
in the sky. As discussed in section~\ref{ssec:obsonez},
our observable~$\DDD(\Vang)$ in \EQR{eq:DDD}
corresponds to the CMB temperature~$T(\Vang)$, and the source 
fluctuation~$\de_g$ is absent, as it is unbiased or we explicitly solve
the temperature evolution using the Boltzmann equation. 
Moreover,  we need to 
pay attention to the fact that the temperature measurements give 
$T(\Vang)=\bar T[1+\Theta(\Vang)]$, or the sum of the background 
temperature~$\bar T$ and its fluctuation~$\Theta(\Vang)$. Therefore, the
full cosmological information contents are described by \EQR{eq:fisher2d}
\beeq
\label{eq:cmbgood}
F_{\mu\nu}={4\pi\over C_0}\left(
{\pa\ln\bar T\over\pa p_\mu}\right)\left({\pa\ln\bar T\over\pa p_\nu}
\right)+\sum_{l=0}
^\infty{2l+1\over2}{\pa\over\pa p_\mu}\bigg(\ln \bar T^2C_l\bigg)
{\pa\over\pa p_\nu}\bigg(\ln\bar T^2C_l\bigg)~,
\eneq
to contrast to the standard Fisher information matrix for CMB
\beeq
\label{eq:cmb}
F_{\mu\nu}^\up{std}=\sum_{l=2}^\infty
{2l+1\over2}{\pa\over\pa p_\mu}\bigg(\ln C_l\bigg)
{\pa\over\pa p_\nu}\bigg(\ln C_l\bigg)~,
\eneq
where we sum the standard Fisher matrix
from the quadrupole $l=2$. The standard Fisher information
matrix can be derived from the full Fisher information, if we assume that
the background CMB temperature~$\bar T$ is precisely known and drops out
of the model parameters, and if we assume that the monopole and the dipole
contain {\it no} cosmological information. However, as we argue, neither
of these  assumptions is correct.

In CMB observations, we obtain the ``background'' CMB temperature 
$\AVE{T}^\up{obs}$ by averaging $T(\Vang)$ over the sky
as defined in \EQR{eq:Tobs}.
As emphasized in \EQR{eq:monozero}, however, the monopole 
fluctuation~$\Theta_0$ is not zero (but note $\Theta^\up{obs}_0=0$),
and hence the observed CMB temperature~$\AVE{T}^\up{obs}$ we use is
{\it not} the background CMB temperature $\bar T$ 
(or $\bar\rho_\gamma$). 
With the Fisher information matrix~$F_{\mu\nu}$ in \EQR{eq:cmbgood},
our estimate of~$\bar T$ is subject to the cosmic
variance set by the monopole~$C_0$ of the power spectrum, and this part is
ignored in the standard analysis $F_{\mu\nu}^\up{std}$. We suspect that
the monopole is ignored, because it is already absorbed in~$\AVE{T}^\up{obs}$.
The observed 
dipole~$\Theta_1$ also contains cosmological information, as it is a
measure of the relative velocity between the observer and the CMB fluid,
all of which can be predicted in a given cosmological model. But the dipole, being mainly due to our local velocity,
is subject to significant nonlinear clustering and galaxy formation which we cannot compute in detail.

We proceed to qualitatively compute the impact on the standard analysis,
ignoring the monopole and the dipole contributions in~$F_{\mu\nu}$.
Given $N_p$-number of cosmological parameters, we need to consider
one extra parameter or the background CMB temperature~$\bar T$, such that
the full parameter analysis contains $N_p+1$ parameters with $p_0:=\ln\bar T$
and the full Fisher information matrix is given as 
\beeq
F_{\mu\nu}
=\left(\begin{array}{cc}F_{00}&F_{0\sigma}\\ F_{\rho0}&F_{\rho\sigma}
\end{array}
\right)~,\Dquad \mu,\nu\in(0,1,\cdots,N_p)~,\qquad 
\rho,\sigma\in(1,\cdots,N_p)~,
\eneq
where $F_{\rho\sigma}\equiv F_{\rho\sigma}^\up{std}$ and
the extra components in addition to the standard Fisher information 
matrix are
\bear
F_{00}&=&{4\pi\over C_0}+\sum_l{2l+1\over2}\left(2+{\pa\ln C_l\over\pa\ln\bar T
}\right)^2~,\\
F_{\rho0}&=&\sum_l{2l+1\over2}\left({\pa\over\pa p_\rho}\ln C_l\right)
\left(2+{\pa\ln C_l\over\pa\ln\bar T}\right)~.
\enar
While the power spectrum~$C_l$ decays exponentially at high~$l$ and the 
summation over~$l$ with the $\bar T$-derivative converges,
these extra components are expected to be large. 
To derive the information loss in the standard CMB analysis,
we need to marginalize over the background CMB temperature.
The full covariance matrix or the inverse of the full Fisher matrix is 
\beeq
F^{-1}_{\mu\nu}=\left[\begin{array}{cc}\left(F_{00}-F_{0\epsilon}F^{-1}_{\epsilon\delta}F_{\delta0}
\right)^{-1}& -{F_{0\epsilon}\over F_{00}}\left(F_{\epsilon\sigma}-F_{\epsilon0}
F_{0\sigma}/F_{00}
\right)^{-1} \\  -{F^{-1}_{\rho\kappa}F_{\kappa0}\over F_{00}-F_{0\epsilon}F^{-1}_{\epsilon\delta}F_{\delta0}}
& \left(F_{\rho\sigma}-F_{\rho0}F_{0\sigma}/F_{00}\right)^{-1}
\end{array}\right]~,\Dquad \epsilon,\delta,\kappa\in(1,\cdots,N_p)~,
\eneq
and after marginalizing over the background CMB temperature~$\bar T$ 
(or $p_0$) we invert the marginalized covariance matrix to obtain the
reduced Fisher information matrix for $N_p$-parameters as
\beeq
\tilde F_{\rho\sigma}=F_{\rho\sigma}^\up{std}-{F_{\rho0}F_{0\sigma}\over F_{00}}~.
\eneq
It is clear that some information is lost, compared to the standard Fisher
matrix $F_{\rho\sigma}^\up{std}$. However, the detailed analysis of the 
information loss and its impact on cosmological parameters will require 
extensive numerical investigations based on the Boltzmann equation solvers,
which is beyond the scope of the present paper. 
We defer this investigation for future work \cite{wedoYves}.

\subsection{3D Weak gravitational lensing and tomography}
\label{ssec:3Dlensing}
3D weak lensing was developed \cite{HEAVE03} for the first time to utilize
the additional redshift information in lensing surveys. Traditionally,
 shape measurements are made for individual galaxies. While their angular
positions are immediately available, the redshift is often unavailable,
and the theoretical predictions are then projected along the line-of-sight 
as in section~\ref{ssec:angobs} to
compare to the observations (see \cite{MELLI99,BASC01,REFRE03,MUVAET08} 
for review). However, even in this case, we need the average radial 
distribution $d\bar N/dzd\Om$
of the source galaxies to correctly compute the theoretical
expectation $\dd^\up{pro}$ in \EQR{eq:theory2d},
and this is often achieved in observations with spectroscopic
or photometric redshift measurements for a subset of the source galaxies.
Furthermore, the recent technological advances allow fairly accurate
photometric redshift measurements for individual galaxies
in lensing surveys (see, e.g., \cite{LSST04,DESC05,EUCLID11,KIDS15}),
and this additional radial information is indeed useful to handle the 
systematic errors in lensing data such as the intrinsic alignments
\cite{HEREHE00,HIMAET04}.

Using this extra information in the radial distribution, a lensing tomography
was proposed \cite{HU99} and is now widely used in lensing surveys
(see, e.g., \cite{KIDS15}), in which the source galaxies are grouped
into several radial 
bins according to their redshifts and weak lensing measurements
are made for individual bins to obtain their auto and cross correlations
among the radial bins. Compared to the traditional weak lensing case,
the covariance can be readily extended, and the orthonormality condition
in the tomographic lensing is then
\beeq
(\CCC~\KKK)_{12}=\de_{z_1z_2}\de^D(\Om_1-\Om_2)d\Om_1~,
\eneq
where $z_i,z_j$ indicate the tomographic bins. There exists an extra structure
for radial bins in the orthonormal relation of  tomographic lensing,
and this is to be contrasted to \EQR{eq:inverseang}
for the traditional weak lensing. The maximum cosmological information
contents can be quantified by the Fisher matrix, which takes the same
form as in the traditional weak lensing, but with the angular power 
spectrum~$C_l^\up{std}$ now replaced with~$C_l^\up{tomo}$
and the inverse angular power spectrum~$\II_l^\up{std}$ with~$\II_l^\up{tomo}$
\beeq
C_l^\up{tomo}
:=\left(\begin{array}{cc}C_l^{11}&C_l^{12}\\  C_l^{12}&C_l^{22} 
\end{array}\right)~,\qquad\qquad
\II_l^\up{tomo}:=\left(C_l^\up{tomo}\right)^{-1}~,
\eneq
where we assumed there are two tomographic bins. This recovers
the original formulation in \cite{HU99}.

3D weak lensing can be viewed as the tomographic weak lensing in the limit,
wheres the number of tomographic bins becomes infinite. In this sense, 
3D weak lensing \cite{HEAVE03} provides the most comprehensive method to
use the full information available in lensing surveys. It 
decomposes the angular position of the shape measurements on the sky
in terms of spherical harmonics and the radial position in terms of 
spherical Bessel function, as described in section~\ref{ssec:3dinv}.
Since more information is used in 3D weak lensing than in the traditional
method or tomographic method,
more cosmological information can be extracted in 3D weak lensing,
and it is important to quantify the net 
increase in the cosmological information.
For illustration, the cosmological information in 3D weak lensing was
approximated \cite{HEAVE03} as the two-dimensional one in \EQR{eq:fisher2dN}
for each Fourier mode to be summed over, and it was found that
$\sim30\%$ improvements can be achieved in measuring the underlying matter
density power spectrum, though the number depends on the characteristics
of the survey.
However, this procedure of computing the information contents
essentially ignores the radial correlation between the observed 
data points, and the correct cosmological information in 3D weak lensing
needs to be computed by using the full three-dimensional Fisher information
matrix in \EQR{eq:fisher3d}, instead of \EQR{eq:fisher2dN}.
The detailed analysis of cosmological information in 3D weak lensing
and tomography will be investigated in future work \cite{wedoNastassia}.

\subsection{Cosmic variance on the baryon density $\bar\rho_b$: Missing 
baryons in the local Universe}
\label{ssec:baryon}
Recent measurements of the cosmic microwave background anisotropies and
the light element abundance from the big bang nucleosynthesis yield a very precise
value for the background baryon density~$\bar\rho_b$ today. While the 
high-redshift measurements of the Lyman alpha forests yields a baryon
density consistent with the background baryon density~$\bar\rho_b(z)$ 
at the corresponding redshift, it is well-known that the baryon density
in the local Universe or low redshift accounts for only about a half 
of~$\bar\rho_b$, and this issue is known as the missing baryon problem 
\cite{FUHOPE98} (see \cite{SHSMDA12} for a recent review).
The baryons at low redshift are expected to be in the warm-hot intergalactic
medium (WHIM), and they are notoriously difficult to observe
in optical or X-ray telescopes (see \cite{NIMAET05,NIKAET18}, 
however, recent measurements of O$_\up{VII}$  in soft X-rays).
In light of the formalism developed in this work, 
we are interested in estimating the cosmic variance on
the observations of the baryon density at low redshift, rather than 
proposing another solution to the missing baryon problem. At low redshift,
the light cone volume is  small and observations are subject to considerable
 cosmic variance. This also applies to the measurements of the baryon
density, as observations of the baryon density~$\rho_b(\xvec)=\bar\rho_b(z)
(1+\de_b)$ include not only the background baryon 
density~$\bar\rho_b(z)$, but also its perturbation~$\de_b$. 

Consider estimating
the baryon density by observing gas clouds (or WHIM), in which the direct
observables are often surrogates for the baryon density such as the oxygen 
number density $\DDD=n_\up{O_{VII}}$ (in a specific ionization state) and
the host galaxy fluctuation in this case becomes the gas cloud fluctuation
$\de_g=\de_\up{WHIM}$. The oxygen abundance needs to be converted to
the overall baryon density, but here we simplify the situation by assuming
that this procedure is straightforward and we use $\DDD=\rho_b(\xvec)$.
Since our interest is in one number or the (background) baryon density~$\bar
\rho_b$ at $z=0$, we can scale out the redshift dependence, such that we 
obtain 
the observational estimate of $\bar\rho_b^\up{obs}$ by averaging  all the
measurements over the light cone volume in similarity to \EQR{eq:dataavg}:
\beeq
\bar\rho_b^\up{obs}\equiv\bar\dd^\up{obs}:={1\over N}\sum_{i=1}^N\dd_i^\up{obs}
(1+z_i)^3~,
\eneq
where $N$ is indeed a few for the case of local baryon density measurements.

This average can be modeled as in \EQRS{eq:bgddd} and~\eqref{eq:perddd}
by using $\bar\DDD(z)=(1+z)^3\bar\rho_b(z)$ and~$\de_g=\de_\up{WHIM}$. While
the full information content in these observations is described
by \EQR{eq:fisher3d}, we  focus on the dominant contribution or the
cosmic variance on the background baryon density~$\bar\rho_b$ in 
\EQR{eq:fisher3dA}. Assuming that the cosmological parameters 
other than $\bar\rho_b$ are known,
and the observations are at low redshift, we first compute
the Fourier kernel in \EQR{eq:GD}
\beeq
\GD_\mu=\int dz~j_0(k\rbar){\pa\over\pa\bar\rho_b}\ln\hdd(z)={1\over
\bar\rho_b}\int dz~j_0(k\rbar)\simeq{\Delta z\over\bar\rho_b}j_0\left({k\Delta
z\over H_0}\right)~,
\eneq
and the Fisher information matrix in \EQR{eq:fisher3dA} becomes
\beeq
F_{\mu\nu}\approx\left({4\pi\Delta z\over\bar\rho_b}\right)^2\int dk \int dk'
~{kk'\over2\pi^2}\IS_0(k,k')j_0\left({k\Delta z\over H_0}\right)
j_0\left({k'\Delta z\over H_0}\right)~,
\eneq
where $\Delta z$ is the survey depth in redshift. Due to the rapid oscillations
and decay of the spherical Bessel function, the integral receives appreciable
contributions only over the $k$-range, where the argument of the spherical
Bessel function is less than about 2,
\beeq
0\leq {k\Delta z\over H_0}\leq 2~.
\eneq
Therefore, the Fisher matrix is
\beeq
F_{\mu\nu}\approx {16H_0^4\over\Delta z^2\bar\rho^2_b}\IS_0(\bar k,\bar k)~,
\eneq
where we used $j^2_0\simeq0.5$ over the range and $\Delta k\simeq2H_0/\Delta z$
and defined the representative wave number $\bar k:=H_0/\Delta z$.

To proceed, we make a series of assumptions to compute the inverse 
spherical power spectrum $\IS_0(k,k')$ by using \EQR{eq:masterIS}. 
As in section~\ref{ssec:lumi},
the gas cloud (or WHIM) is dominated by the matter density clustering
$\de_g\simeq\de_m$, and the spherical power spectrum of the matter density
fluctuation at low redshift is isotropic $S_l(k,k')\approx\de^D(k-k')P_m(k)$
(see Appendix~\ref{app:ob}).
The Fourier angular kernel in \EQR{eq:FF} is non-vanishing 
\beeq
\FF_l(k_1,k_2):=\int dz~j_l(k_1\rbar)j_l(k_2\rbar)\approx \frac12\Delta z~,
\eneq
only over $0\leq k_1,k_2\leq\Delta k$, and the integral 
equation~\eqref{eq:masterIS} becomes
\beeq
1\approx{8\over\pi^2}{H_0^7\over\Delta z^5}\IS_0(\bar k,\bar k)P_m(\bar k)
\simeq{16H_0^4\over\Delta z^2}\IS_0(\bar k,\bar k)\sigma^2_R~,
\eneq
where we defined the rms matter fluctuation smoothed by a top-hat 
radius~$R=\Delta z/H_0$:
\beeq
\sigma^2_R:=\int d\ln k~{k^3\over2\pi^2}P_m(k)j_0^2(kR)
\simeq\left({H_0\over\Delta z}\right)^3{P_m(\bar k)\over2\pi^2}~.
\eneq
The full Fisher matrix simplifies accordingly, and
the uncertainty on the baryon density becomes 
\beeq
F_{\mu\nu}^{-1/2}\approx \bar\rho_b\sigma_R~.
\eneq
The cosmic variance is driven by the matter density fluctuations, and at low
redshift it is similar to the rms fluctuation smoothed with the scale set by
the redshift depth. With $\Delta z\simeq0.1$, the comoving radius is about
$R\simeq300~\hmpc$, and the rms fluctuation $\sigma_R\simeq0.06$. 
The rms fluctuation further decreases to $\sigma_R\simeq0.002$,
as the survey depth increases to $\Delta z\simeq0.3$.

Given the measurement uncertainties and the amount of missing baryons
in the local Universe, the cosmic variance contributes only a small fraction 
to the problem. However, it is important to know that the missing 50\% correspond to about $8.3\sigma_R$. Furthermore, our simple estimate is based on many simplifying
assumptions: First, we computed the maximum cosmological information
in an idealized survey, where an infinite number of measurements can be
made. However, real observations of the local baryon density are indeed
based on a few sight lines towards bright background sources, dramatically
increasing the sample variance in real observations, 
compared to our cosmic variance limit.
Second, the location of the observers is not a random place in the Universe, 
but a highly biased placed (or a Milky-way sized halo). Furthermore, while
we used the linear theory to compute the cosmic variance, the real analysis
has to account for the nonlinear effects of galaxy clustering.
These two effects will greatly increase the variance in the local baryon
density measurements. The detailed analysis of all these effects
will require numerical simulations, and it will  be investigated in future work
\cite{wedoBaryon}.

\subsection{Galaxy power spectrum in a cube vs spherical power spectrum
on a light cone}
\label{ssec:gc}
The standard galaxy power spectrum analysis
proceeds as though the survey volume is in the hypersurface of simultaneity
and a Fourier decomposition is made in the rectangular box.
The uncertainties of the power spectrum estimates
 are further reduced by the number of Fourier
 modes available in the survey volume. This is qualitatively correct, 
if the survey volume is small enough,
but it becomes inaccurate as we look into the large scale
modes and the survey volume becomes larger. Here we make the connection
of this traditional power spectrum analysis to our spherical power spectrum
analysis, by which we quantify what conditions are needed to justify
the small volume.

We first need to compute the inverse spherical power spectrum~$\IS_l(k,k')$
given in \EQR{eq:masterIS}. Under the assumption that the sky coverage is
small and the redshift depth is shallow, 
we will make use of a series of manipulations based on the 
rapid oscillating properties of the spherical Bessel function, called the Limber approximation~\cite{LIMBE54,LOAF08}
\beeq
\label{eq:idt}
\int_0^\infty dx~x^2f(x)j_l(\alpha x)j_l(\beta x)\simeq{\pi\over2\alpha^2}
\de^D(\alpha-\beta)f\left(x=l+\frac12\right)~,
\eneq
where the function~$f(x)$ is assumed to be slowly varying over the range
relevant to the integration and it becomes the identity when $f(x)$ is
a constant. Assuming also $S_l(k,k')\approx\de^D(k-k')P_m(k)$ and
performing the integration over~$k_1$, we obtain
\beeq
\left({2\over\pi}\right)\int dk_1'~k_1'k_B\IS_l(k_1',k_A)\times
\int dr~{H^2\over\rbar^2}j_l(k_1'\rbar)j_l(k_B\rbar)P(k_\star)=\de^D(k_A-k_B)~,
\eneq
where the Hubble parameter was introduced by converting the integration
variable from~$dz$ to~$d\rbar$ and the star indicates that the Fourier
mode is evaluated under the condition $k_\star\rbar=l+1/2$.
Further assuming that the inverse spherical power spectrum is
$\IS_l(k,k')\approx\de^D(k-k')\IS_l(k)$, and integrating over~$k_A$,
the closed equation can be manipulated as
\beeq
1\simeq\int d\rbar~{H^2\over\rbar^2}P(k_\star)\times{2\over\pi}
\int dk_A~k_A^2\IS_l(k_A)j_l(k_A\rbar)j_l(k_A\rbar)~.
\eneq
Applying the same trick for the spherical Bessel function one
more time to the integration over~$\rbar$ and simplifying the remaining
integral with the Dirac delta function, 
we derive the inverse spherical power spectrum
\beeq
\label{eq:invSL}
\IS_l(k)\simeq\left({\rbar^2\over H}\right)^2_\star P^{-1}(k)~,
\eneq
where the star indicates now that the comoving radius is evaluated 
under the condition $k\rbar_\star=l+1/2$. For a small survey volume, 
where the flat-sky approximation is accurate and the redshift evolution
is negligible, the inverse spherical power spectrum is literally 
the inverse of the  power
spectrum, but with the volume factor to compensate for the dimensionful
quantity.

Having derived the inverse spherical power spectrum, we are now in a position
to tackle the more complicated equation for the full Fisher information
matrix in \EQR{eq:fisher3d}. Under the same assumption that the survey
volume is small, we can ignore the Fourier angular kernel
\beeq
\NK_{l,\mu}(k_1,k_2)=\int dz~j_l(k_1\rbar){\pa\over \pa p_\mu}j_l(k_2\rbar)
\simeq0~.
\eneq
In addition, we assume that the background galaxy number density 
$\hdd:=\bar n(z)$ is known, so that we can ignore the other Fourier 
angular kernel
\beeq
\HK_{l,\mu}(k_1,k_2)=\int dz~j_l(k_1\rbar)j_l(k_2\rbar){\pa\over\pa p_\mu}
\ln\bar n(z)\simeq0~,
\eneq
though we can only measure $\bar n(z)$ up to the monopole contribution.
This assumption also eliminates the mean contribution to the Fisher matrix,
and only the covariance of the galaxy number counts contributes to the
Fisher matrix. Furthermore, we only consider the monopole power 
spectrum $S_0(k,k')$ to make a connection to the angle-averaged power spectrum.
The traditional power spectrum analysis proceeds
as if the (small) survey volume is embedded in a cubic volume of hypersurface
with the origin at the center of the cubic volume, while the observer
is indeed at a distance, so that the line-of-sight direction is considered
 fixed over the survey volume. Under this assumption, the observed
power spectrum is well approximated as the redshift-space power spectrum
described by the Kaiser formula \cite{KAISE87}, and the redshift-space
power spectrum has the monopole, the quadrupole and the hexadecapole only.
The information contents for this monopole power spectrum were derived
\cite{FEKAPE94,TEGMA97}:
\beeq
F_{\mu\nu}^\up{std}=2\pi\int d\ln k\left({k\over2\pi}\right)^3V_\up{eff}
\left[{\pa\over \pa p_\mu}\ln P(k)\right]\left[{\pa\over \pa p_\nu}\ln P(k)
\right]~,
\eneq
where $P(k)$ represents the monopole power spectrum and
\beeq
V_\up{eff}:=\int d^3x\left[{\bar n(x)P(k)\over1+\bar n(x)P(k)}\right]^2~
\eneq
is the survey volume in our idealized case ($\bar n\RA\infty$).

This monopole power spectrum or the angle-average
of the power spectrum in Fourier space can be considered as the observed
angle-average of the power spectrum if the observer is located at the center
of the survey volume. Under this assumption, the monopole power spectrum
corresponds to our monopole spherical power spectrum~$S_0(k,k')$. 
Therefore, these assumptions greatly simplify the Fisher information matrix in
 \EQR{eq:fisher3d} to
\beeq
F_{\mu\nu}\simeq {(4\pi)^4\over2}
\left(\int d\ln k_1 {k_1^3\over2\pi^2}\cdots\int d\ln k_4 {k_4^3\over2\pi^2}
\right)\IS_0(k_1)\IS_0(k_3)
\FF_0^{12}\FF_0^{23}\FF_0^{34}\FF_0^{41}
{\pa\over\pa p_\mu}P(k_2){\pa\over\pa p_\nu}P(k_4) ~.
\eneq
Our strategy is again to apply the Limber approximation \EQR{eq:idt} multiple
times to simplify the integration over the wavevector and redshift.
We first perform the integration over~$k_1$ and~$k_3$ with \EQR{eq:idt}
and simplify the Dirac delta function to derive
\bear
F_{\mu\nu}&\simeq&{(4\pi)^2\over2}\int d\ln k_2~{k_2^3\over2\pi^2}
\int d\ln k_4~{k_4^3\over2\pi^2}\int d\rbar_i\left({H_i\over\rbar_i}\right)^2
\int d\rbar_j\left({H_j\over\rbar_j}\right)^2    \\
&&\times
j_0(k_2\rbar_i)j_0(k_2\rbar_j)j_0(k_4\rbar_j)j_0(k_4\rbar_i)
\IS_0\left(k_1={1\over2\rbar_i}\right)\IS_0\left(k_3={1\over2\rbar_j}
\right){\pa\over\pa p_\mu}P(k_2){\pa\over\pa p_\nu}P(k_4) ~.\nn
\enar
Applying \EQR{eq:idt} to the integration over~$k_4$ and
using the expression for the inverse spherical power spectrum~$\IS_0(k)$
in \EQR{eq:invSL}, we obtain
\beeq
F_{\mu\nu}\simeq2\pi\int d\ln k~{k^3\over2\pi^2}{\pa\over\pa p_\mu}P(k)
\int d\rbar~\rbar^2j_0(k\rbar)j_0(k\rbar)P^{-1}\left(k_\star\right)
{\pa\over\pa p_\nu}\ln P(k_\star) ~,
\eneq
where $k_\star:=1/2\rbar$. The radial integral is then re-arranged 
by using \EQR{eq:idt} to arrive
at the desired equation of the standard power spectrum analysis
\beeq
F_{\mu\nu}\simeq2\pi\int d\ln k~\left({k\over2\pi}\right)^3V_\up{eff}
\left[{\pa\over\pa p_\mu}\ln P(k)\right]\left[
{\pa\over\pa p_\nu}\ln P(k)\right] ~,
\eneq
where the effective volume is 
\beeq
V_\up{eff}:=4\pi\int d\rbar~\rbar^2j_0^2(k\rbar)\simeq{2\pi R\over k^2}~,
\eneq
where $R$ denotes the survey depth and we ignored the lower boundary
of the survey.
Compared to the spherical Fourier analysis, 
the standard power spectrum analysis in summary makes a series of
approximations:
(1)~the redshift evolution over the survey volume is negligible,
(2)~the angular position is constant (distant-observer approximation),
(3)~the angular diameter distances~$\rbar$ are independent of cosmological
parameters ($\NK_{l,\mu}=0$), (4)~the background galaxy number density is 
known. 

It is well known that the redshift-space power spectrum contains
more information than just the monopole power spectrum. It is evident
now that a lot more cosmological information is available in galaxy surveys
and not all the information has been utilized in the traditional power spectrum
analysis. The detailed power spectrum analysis will be investigated
in future work \cite{wedo2}.

\section{Discussion and Summary}
\label{sec:discuss}
In this paper, we have developed a theoretical framework to describe 
cosmological observables on the light cone 
and we have  derived the Fisher information matrix to quantify the maximum cosmological
 information obtainable from  cosmological observables such as the
luminosity distance, weak gravitational lensing, galaxy clustering, and the
cosmic microwave background (CMB) anisotropies. 
As all the cosmological
observables contain perturbations, their measurements are subject to the 
cosmic variance, and in computing the cosmic variance, we have taken into
account  that the survey geometry is  the light cone volume.
In section~\ref{sec:app},
we have discussed in detail the impact of our formulation on
the cosmological information contents for five different
cosmological observables, in comparison to the standard analysis.
Our main findings are as follows:

\begin{itemize}
\item

Our theoretical framework provides a unified description of  angular
observables,  observables with  redshift information, and their variants
such as the projected observables.
Moreover, it accounts for the fact that  observables are often
obtained with weights given by the number counts of  host galaxies.
The measurements of type Ia supernovae
are, for instance, modulated not only by the fluctuations in the luminosity 
distance itself, but also by the spatial correlation of the host galaxies, 
as we can only have supernovae in a  host galaxy.
While the latter is often ignored in literature, it is indeed the  dominant source of 
perturbations.

\item
To properly quantify the cosmological information contents  that can be 
derived from
a given  observable, we have deployed the
Fisher information technique and assumed a Gaussian probability distribution.
As the cosmological observables can be measured
over a range of redshift, we need to account for their three-dimensional
correlation on the light cone and to derive its inverse in computing the
Fisher information matrix. In the standard picture, where our survey volume
is treated as a cubic box, the Fourier analysis provides the best way for
this task, and the inverse of the power spectrum is trivial, as
each Fourier mode is independent. However, in the real Universe, where 
the survey volume is on the past light cone, the radial and the angular 
positions carry different information. To properly accommodate this,
we have used the spherical Fourier analysis and derived for the first time
the closed equation~\eqref{eq:masterIS} 
for the inverse of the three-dimensional correlation 
of the cosmological observables.  We have fully taken into account that the lightcone geometry breaks translation invariance in the
radial direction and therefore radial Fourier modes are correlated. 

\item
Given the inverse spherical power spectrum, it is straightforward to 
derive the Fisher information matrix for 
 three-dimensional cosmological observables. To obtain rough estimates
for the impact of  cosmic variance, we have applied it to 
supernova observations,  local baryon density measurements, 
3D weak gravitational lensing, and galaxy clustering, 
all of which are three-dimensional on the light cone and are 
correlated (see section~\ref{sec:app} for detailed discussions).
While the first two are often 
thought to provide  measurements of 
background quantities such as the background luminosity distance and the
global baryon density, they both measure only the sum of the background and 
the perturbation together, and hence these measurements are also subject to 
cosmic variance (sections~\ref{ssec:lumi} and~\ref{ssec:baryon}).

\item
For  three-dimensional cosmological probes such as galaxy clustering and
3D weak lensing, the spherical Fourier analysis is used to
analyze the observables on the light cone, and our Fisher matrix analysis
in section~\ref{ssec:gc}
shows that the standard analysis is based on many simplifying assumptions
such as the distant observer, the flat-sky, and no radial correlations.

\item
Regarding angular cosmological observables such as  CMB
anisotropies and  weak lensing observables, the standard
formalism correctly describes these angular observables on the light cone,
except one subtlety associated with the observed mean in the CMB temperature.
The observed mean of the CMB temperature is  obtained by averaging
the CMB temperature on the sky, which includes not only the background~$\bar T$
(an input cosmological parameter), but also the monopole perturbation
(a prediction of the model). While the cosmic variance in the observed temperature
is expected to be $10^{-5}$, its present measurement 
error~\cite{FICHET96,FIXSE09} is about $2.1\times 10^{-4}$ which is similar
to the error in e.g. the angular scale subtended by the acoustic peaks~\cite{PLANCKcos18} which is $\Delta\theta_* = 3\times 10^{-4}$. This error propagates to the power spectrum measurements
into cosmological parameter estimation. A proper analysis of this subtlety in CMB observations
will quantify its impact on the cosmological parameter analysis 
\cite{wedoYves}.
\end{itemize}

In this paper, we derived the maximum cosmological information contents
from a cosmological observable. This cosmic variance limit arises, because
we have a single light-cone volume at our disposition for observations. 
A way to make maximal use of it
is simply to measure the cosmological observables up to higher redshift, 
at which
the light cone volume is large enough to overcome the disadvantage from a
single observation point.  A more practical solution
is the multi-tracer method \cite{SELJA09,MCSE09}, already
developed in the standard analysis,  and it can be easily 
generalized to the light 
cone analysis. The idea there is to consider  several different cosmological
observables  which trace the same underlying density field, 
so that by measuring
those observables, one can effectively increase the sampling rate and 
may eliminate the stochasticity completely for certain cosmological parameters
in an idealized case.

\acknowledgments
We acknowledge useful discussions with Avi Loeb and Matias Zaldarriaga.
We acknowledge support by the Swiss National Science Foundation.
J.Y. and E.M. are further supported by
a Consolidator Grant of the European Research Council (ERC-2015-CoG grant
680886).

\appendix
\section{Spherical Fourier analysis with the 
observed redshift as a dimensionless radial distance}
\label{app:sp}

We have developed the spherical Fourier decomposition of the cosmological
observables in section~\ref{ssec:3dinv} and computed the Fisher information 
matrix in a light cone volume in section~\ref{ssec:infovol}. In this formalism,
the cosmological observable on a light cone can be naturally decomposed.
However, there exists one subtlety, albeit not a problem:
The Fourier analysis of the radial modes is based on the
comoving distance~$\rbar$ and its Fourier mode~$k$:
\beeq
\xvec=\rbar_z\Vang~,\Dquad k\sim1/\rbar_z~,\qquad [k]=L^{-1}~,
\eneq
and the conversion of the observed redshift~$z$ into the comoving 
distance~$\rbar_z$ involves a cosmological model. Here we present an 
alternative method to perform the spherical Fourier analysis, in which we
construct a new observer coordinate and its Fourier counterpart:
\beeq
\xvec=z\Vang~,\Dquad k\sim1/z~,\qquad [k]=1~.
\eneq
This is rather unconventional, but we can readily apprehend its advantage:
The spherical power spectrum can be constructed out of the raw observed 
data in a model independent way. A slight disadvantage arises when we
compare to the standard theoretical predictions. For example, we understand
well how much power is at $k=1\hmpci$, while we will need a model-dependent
conversion to understand the power at the dimensionless scale
$k=1$. A simple calculation shows that
$k=1~\hmpci$ would correspond
to $\rbar\approx2\pi~\hmpc$, which would then correspond to the redshift
$z\approx0.002$ in a $\Lambda$CDM with $\Omega_m=0.3$, 
hence the dimensionless
Fourier number would be $k\approx2\pi/0.002\simeq3100$.
Though this change amounts to a simple conversion of units at low redshift,
the relation between the comoving distance and the redshift is highly 
nonlinear at high redshift, and the comparison to the standard analysis 
becomes non-trivial.

The spherical Fourier analysis proceeds almost exactly the same way in
section~\ref{ssec:3dinv} with the comoving distance~$\rbar$ replaced by
the redshift~$z$. The cosmological observables are now decomposed as
\bear
\Ddd(\xvec)&:=&\sum_{lm}\int_0^\infty dk\sqrt{2\over\pi}~kj_l(kz)Y_{lm}
(\Vang)\sss_{lm}(k)~,\\
\sss_{lm}(k)&\equiv&
\int d\Omega\int dz~z^2\sqrt{2\over\pi}~kj_l(kz)Y_{lm}^*(\Vang)\Ddd(\xvec)~.
\enar
The Fourier mode~$\sss_{lm}(k)$ and its spherical power spectrum~$S_l(k,k')$
are now dimensionless. For the two-point~$\xi_{12}$ and its 
inverse~$\zeta_{12}$ correlation functions, the decomposition remains
almost unchanged as in \EQRS{eq:xi3d} and~\eqref{eq:zetaFT}
\bear
\xi_{12}&=&4\pi\sum_{lm}\int dk\int dk'~{kk'\over2\pi^2}S_l(k,k')
j_l(kz_1)j_l(k'z_2)Y_{lm}(\Vang_1)Y_{lm}^*(\Vang_2)~,\\
\zeta_{12}&=:&4\pi\sum_{lm}\int dk\int dk'~{kk'\over2\pi^2}\IS_l(k,k')
j_l(kz_1)j_l(k'z_2)Y_{lm}(\Vang_1)Y_{lm}^*(\Vang_2)~,
\enar
which also defines the ``inverse'' spherical power spectrum~$\IS_l(k,k')$. 
The relation between the spherical power spectrum and its inverse
is exactly the same as in \EQR{eq:masterIS}, though
the Fourier angular kernel is slightly different
\beeq
\FF_l(k_1,k_2):=\int dz~j_l(k_1z)j_l(k_2z)={\pi\over2(2l+1)}{(k_<)^l
\over(k_>)^{l+1}}~,
\eneq
yet analytically solvable, if the integration is performed from zero to
infinity, where $k_>$ is the maximum of 
$k_1$ and $k_2$. Despite this analytic solution, 
the closed equation for the inverse spherical power spectrum 
cannot be further simplified, as the spherical power spectrum is an unknown
input function. The inverse covariance in the light cone volume is 
\beeq
\KKK_{12}={4\pi\over\hdd(z_1)\hdd(z_2)}
\sum_l{2l+1\over4\pi}P_l(\ga_{12})\int dk\int dk'~{kk'\over2\pi^2}\IS_l(k,k')
j_l(kz_1)j_l(k'z_2)~.
\eneq

We now proceed to compute the Fisher information matrix in a light cone
volume in section~\ref{ssec:infovol}. The contribution of the mean to
the Fisher information matrix is exactly the same as
\beeq
\frac12\Tr\bigg[\KKK\MMM_{\mu\nu}\bigg]=
(4\pi)^2\int dk\int dk'~{kk'\over2\pi^2}\IS_0(k,k')\GD_\mu(k)\GD_\nu(k')~,
\eneq
with the same Fourier kernel
\beeq
\GD_\mu(k):=\int dz~j_0(kz){\pa\over\pa p_\mu}\ln\hdd(z)~.  
\eneq
The variation of the covariance in the Fisher information matrix can be
readily obtained simply by replacing~$\rbar$ with~$z$,
and the expression is almost identical to \EQR{eq:inter}, except that
the spherical Bessel functions are mow independent of cosmological parameters
and they can be pulled out of the derivative with respect to the model 
parameters. Consequently, the contribution of the covariance to the
Fisher information matrix has the same structure with the same Fourier
angular kernels:
\bear
&&\frac12\Tr\left[\left(\KKK{\pa\over\pa p_\mu}\CCC\right)\left(\KKK
{\pa\over\pa p_\nu}\CCC\right)\right]=\left({2\over\pi}\right)^4
\sum_l{2l+1\over2}
\left(\prod_{i=1}^4\int dk_i~k_i\right)\left(\prod_{j=1}^4\int dk_j'~k_j'
\right)\IS_l(k_1,k_2)\IS_l(k_3,k_4)
\nnn
&&\qquad\times
\left[\FF_l^{21'}\HK_{l,\mu}^{32'}S_l(k_1',k_2')+
\FF_l^{32'}\HK_{l,\mu}^{21'}S_l(k_1',k_2')
+\FF_l^{21'}\FF_l^{32'}{\pa\over\pa p_\mu}S_l(k_1',k_2')\right]\nnn
&&\qquad\times
\left[\FF_l^{43'}\HK_{l,\nu}^{14'}S_l(k_3',k_4')+
\FF_l^{14'}\HK_{l,\nu}^{43'}S_l(k_3',k_4')
+\FF_l^{43'}\FF_l^{14'}{\pa\over\pa p_\nu}S_l(k_3',k_4')\right]~.
\enar
except that one Fourier kernel is identically vanishing:
\beeq
\NK_{l,\mu}(k_1,k_2):=\int dz~j_l(k_1z){\pa\over\pa p_\mu}j_l(k_2z)\equiv0~.
\eneq

\section{Spherical power spectrum on the light cone}
\label{app:ob}
Fourier analysis and the power spectrum provide the best
way to characterize the initial conditions and their subsequent evolutions
at the linear order in perturbations in a hypersurface of simultaneity.
However, in observations all the cosmological observables are measured
along the past light cone, piercing through different hypersurfaces.
Since the Fourier transformation is intrinsically non-local, the Fourier
analysis of the cosmological observables, 
either traditional or spherical, involves
non-trivial complications due to the time evolution of perturbations.

To illustrate the point, we assume that our cosmological observable is
simply the matter density field $\Ddd(\xvec):=\de_m(\xvec)$ located
at the observed redshift and angle.
The spherical Fourier analysis starts with the decomposition 
in \EQR{eq:3dFT}:
\beeq
\de_m^\up{obs}
(\xvec):=\sum_{lm}\int_0^\infty dk\sqrt{2\over\pi}~kj_l(k\rbar_z)Y_{lm}
(\Vang)\sss_{lm}^\up{th}(k;t_z)~,\Dquad \xvec=(z,\Vang)~,
\eneq
where we used the superscripts~``th'' and~``obs'' 
to indicate that the left-hand side is obtained in observations and the 
right-hand side is our theoretical modeling of the observation.
The Fourier component $\de^\up{th}_m(\bm{k};t)$
of the matter density, for instance, is a good example of the theoretical
quantity, and it is characterized by its power spectrum $P_m^\up{th}(k;t)$,
where the superscripts are often omitted in literature. However, note that
these theoretical quantities are computed in a hypersurface, so that
the Fourier component of the matter density, 
for example, implicitly assumes the time-dependence
$\de_m^\up{th}(\bm{k};t)$, i.e., Fourier decomposition is
performed in a hypersurface of constant~$t$, inaccessible to the observer.
Hence, the spherical Fourier
component~$\sss_{lm}^\up{th}(k;t_z)$ has also
the time-dependence, where the
observed redshift~$z$ specifies the hypersurface.

Completely independent of our theoretical modeling,
the (observed) spherical Fourier coefficients can be obtained
in terms of the (observed) density field~$\de_m^\up{obs}(\xvec)$ as
\beeq
\sss^\up{obs}_{lm}(k)
:=\int d\Omega\int d\rbar~\rbar^2\sqrt{2\over\pi}~kj_l(k\rbar)
Y_{lm}^*(\Vang)\de^\up{obs}_m(\xvec)~,
\eneq
where $\sss_{lm}^\up{obs}(k)$ is independent of time as the time-dependence
is integrated out.
Using the spherical Fourier decomposition above, we derive the
relation between $\sss_{lm}^\up{obs}$ and $\sss_{lm}^\up{th}$ as
\beeq
\label{eq:obft}
\sss^\up{obs}_{lm}(k)
=\int dk'~\left[{2kk'\over\pi}
\int d\rbar~\rbar^2j_l(k\rbar)j_l(k'\rbar)\right]\times\sss_{lm}^\up{th}(k';t_z)
~.
\eneq
Were it not for the light-cone observation (or the time-dependence)
and the finite survey volume,
$\sss_{lm}^\up{th}$ could be pulled out of the line-of-sight
integration, and the relation would simply indicate
\beeq
\sss^\up{obs}_{lm}(k)\equiv\sss_{lm}^\up{th}(k)~,
\eneq
as desired. Due to the time evolution, however, a non-trivial complication
arises for their relation in \EQR{eq:obft}. 

Similarly, the (observed) spherical
power spectrum can be obtained by considering the ensemble average
of the (observed) spherical Fourier components 
\bear
\AVE{s_{lm}^\up{obs}(k)s_{l'm'}^\up{obs^*}(k')}&=&{2\over\pi}
\int d\Omega_1\int d\Omega_2\int d\rbar_1\int d\rbar_2
~\rbar_1^2\rbar^2_2~kk'j_l(k\rbar_1)j_{l'}(k'\rbar_2)\nnn
&&\times
Y_{lm}^*(\Vang_1)Y_{l'm'}(\Vang_2)
\AVE{\de^\up{obs}_m(\xvec_1)\de^\up{obs}_m(\xvec_2)}~.
\enar
The (observed) two-point correlation function is then related to the
theoretical power spectrum:
\beeq
\AVE{\de^\up{obs}_m(\xvec_1)\de^\up{obs}_m(\xvec_2)}=\int{d^3k\over(2\pi)^3}
~e^{i\bdv{k}\cdot(\xvec_1-\xvec_2)}P^\up{th}_m(k;z_1,z_2)~,
\eneq
where the power spectrum involves two different hypersurfaces specified 
by~$z_1$ and~$z_2$ and in linear theory this can be factored out by using
the growth factor~$D(z)$ normalized at some initial time~$t_0$ as
\beeq
P_m^\up{th}(k;z_1,z_2)=D(z_1)D(z_2)P_m^\up{th}(k;t_0)~.
\eneq
Expanding the exponential factor, the ensemble average can be arranged as
\bear
\AVE{s_{lm}^\up{obs}(k)s_{l'm'}^\up{obs^*}(k')}&=&\de_{ll'}\de_{mm'}
\int d\tilde k
\left[{2k\tilde k\over\pi}\int d\rbar_1~\rbar_1^2
j_l(k\rbar_1)j_l(\tilde k\rbar_1)\right]\nnn
&&\times
\left[{2k'\tilde k\over\pi}\int d\rbar_2~\rbar_2^2
j_l(k'\rbar_2)j_l(\tilde k\rbar_2)\right]P_m^\up{th}(\tilde k;z_1,z_2)~,
\enar
and the (observed) spherical power spectrum is therefore
\beeq
S_l^\up{obs}(k,k')\equiv\int d\tilde k
\left[{2k\tilde k\over\pi}\int d\rbar_1~\rbar_1^2
j_l(k\rbar_1)j_l(\tilde k\rbar_1)\right]
\left[{2k'\tilde k\over\pi}\int d\rbar_2~\rbar_2^2
j_l(k'\rbar_2)j_l(\tilde k\rbar_2)\right]
P_m^\up{th}(\tilde k;z_1,z_2)~.
\eneq
Again, in the absence of the time-evolution along the light cone
in a survey with infinite volume, the power spectrum $P_m^\up{th}$
could be pulled out of the line-of-sight integrations, 
and the square brackets simplify to the Dirac delta functions, yielding
\beeq
S_l^\up{obs}(k,k')=\de^D(k-k')P_m^\up{th}(k)~.
\eneq
This simplification is again possible, only if the time-evolution along the
light cone is neglected.
In linear theory, the (observed) spherical power spectrum can be further
simplified by using the growth factor as
\beeq
S_l^\up{obs}(k,k')=\int d\tilde k~P_m^\up{th}(\tilde k;t_0)
\mathcal{T}_l(k,\tilde k)\mathcal{T}_l(k',\tilde k)
\eneq
where we defined a Fourier kernel
\beeq
\mathcal{T}_l(k,\tilde k):=
{2k\tilde k\over\pi}\int d\rbar~\rbar^2j_l(k\rbar)j_l(\tilde k\rbar)D(z)~.
\eneq

\bibliography{ms.bbl}

\end{document}